\documentstyle[11pt,aaspp4,psfig]{article}

\def\gta{\ifmmode {\mathbin{\lower 3pt\hbox   
    {$\,\rlap{\raise 5pt\hbox{$\char'076$}}\mathchar"7218\,$}}}
    \else {${\mathbin{\lower 3pt\hbox
    {$\rlap{\raise 5pt\hbox{$\char'076$}}\mathchar"7218\,$}}}
    $}\fi}
\def\lta{\ifmmode {\,\mathbin{\lower 3pt\hbox   
    {$\,\rlap{\raise 5pt\hbox{$\char'074$}}\mathchar"7218\,$}}}
    \else {${\mathbin{\lower 3pt\hbox
    {$\rlap{\raise 5pt\hbox{$\char'074$}}\mathchar"7218\,$}}}
    $}\fi}

\begin{document}

\title{Intermediate-Mass Black Holes}

\author{M. Coleman Miller$^1$ and E. J. M. Colbert$^2$}
\affil{$^1$Department of Astronomy, University of Maryland\\
       College Park, MD  20742-2421\\
       miller@astro.umd.edu\\
       $^2$Department of Physics and Astronomy, Johns Hopkins University, \\
       Baltimore, MD 21218\\
       colbert@pha.jhu.edu}

\begin{abstract}

The mathematical simplicity of black holes, combined with their
links to some of the most energetic events in the universe, means
that black holes are key objects for fundamental physics and 
astrophysics.  Until recently, it was generally believed that
black holes in nature appear in two broad mass ranges:
stellar-mass ($M\sim 3-20\,M_\odot$), which are produced by the
core collapse of massive stars, and supermassive ($M\sim 10^6-10^{10}
\,M_\odot$), which are found in the centers of galaxies and are
produced by a still uncertain combination of processes.  In the last
few years, however, evidence has accumulated for an intermediate-mass
class of black holes, with $M\sim 10^2-10^4\,M_\odot$.  If such objects
exist they have important implications for the dynamics of stellar
clusters, the formation of supermassive black holes, and the production
and detection of gravitational waves.  We review the evidence for
intermediate-mass black holes and discuss future observational and
theoretical work that will help clarify numerous outstanding questions
about these objects.

\end{abstract}


\section{Introduction}

Isolated black holes are the simplest macroscopic objects in nature,
being completely described by just their gravitational mass, angular
momentum, and electric charge, and only the mass and angular momentum
are likely to be significant for any real black hole.  As a result,
the mathematical theory of black holes has been developed extensively
(e.g., Chandrasekhar 1992) and properties of black holes such as Hawking
radiation (Hawking 1976) are now being compared with predictions of 
string theory (e.g., Maldacena, Strominger, \& Witten 1997).

Direct evidence for the existence of black holes has been slower to
accumulate.  Starting in the early 1970s with the discovery of a number
of bright X-ray sources using the {\it Uhuru} satellite (for a summary
of this mission and sources see Forman et al. 1978), it has
become progressively clearer that there are many binaries in our
Galaxy and others that consist of a black hole of mass $\sim
3-20\,M_\odot$ accreting matter from a stellar companion. Black holes in
this mass range, called stellar-mass black holes, are thought to have
been born during the core collapse of a massive star.  Convincing
evidence that an object is a black hole requires that its mass be
definitely established to be in excess of $3\,M_\odot$, which is an
extremely conservative upper limit to the mass of a neutron star
(e.g., Kalogera \& Baym 1996).
Such mass estimates require careful radial velocity measurements of
the companion star.  At this time, 17 X-ray emitting compact objects
are known to have masses in excess of the neutron star maximum (Orosz 2002).
In some cases the orientation of the orbit and nature of the companion
allow more precise estimates of the mass.  These mass estimates currently
range from $\sim 4\,M_\odot$ (GRO~0422+32) to $14\pm 4\,M_\odot$ 
(GRS~1915+105; see Orosz 2002 for an updated table).

Independent evidence for the existence of supermassive black holes in the
centers of galaxies has also become compelling.  The tremendous
luminosities and small sizes of active galactic nuclei (AGN) led early on
to suggestions that these are powered by accretion onto black holes
(Zel'dovich \& Novikov 1964; Salpeter 1964). This model is now solidly
established by many observations.  Not long after the discovery of
quasars, Lynden-Bell (1969) realized that many ``dead" quasars would exist
as supermassive black holes in nearby galaxies.  This has now been
confirmed by high-precision monitoring of stars. Multiple stellar orbits
have been tracked around the $\sim 3\times 10^6 \,M_\odot$ black hole in
the center of our Galaxy (e.g., Eckart \& Genzel 1996; Ghez et al. 1998;
Ghez et al. 2000;  Eckart et al. 2002; Sch\"odel et al. 2002; Ghez et al.
2003a,b), following up earlier observations of gas motion in the Galactic
center that suggested a strong concentration of mass (e.g., Lacy et al.
1980; Genzel et al. 1985).  Intriguingly, some of the specifics of the
stellar distribution very near the Galactic center may be best explained
by the presence of an intermediate-mass black hole (Hansen \&
Milosavljevic 2003).  Recent observations of the pericenter passage of
star S2 (Sch\"odel et al. 2002) demonstrate that the mass is contained
within a radius of $6\times 10^{-4}$~pc, far more compact than possible
for a stable distribution of individual objects. The spectral energy
distribution of AGN can extend well into the gamma ray regime, implying
relativistic motion.  Finally, relativistically broadened Fe K$\alpha$
lines have been seen from several AGN, and their detailed properties may
suggest rapid rotation as well as  confirming the deep potential well
around black holes (e.g., Iwasawa et al. 1996; Dabrowski et al. 1997;
Wilms et al. 2001; Fabian et al. 2001).

The formation of supermassive black holes is not as well established as
the formation of stellar-mass black holes.  The high luminosity of many
AGN indicates that they are obtaining matter from an accretion disk, and
it is possible that this is their primary mode of growth.  However, it is
also possible that dynamical interactions or relativistic instabilities
could contribute to the growth of supermassive black holes (e.g.,
Begelman, Blandford, \& Rees 1984). Interestingly, it has recently been
established that the mass of a supermassive black hole is tightly
correlated with the velocity dispersion of the stars in the host galaxy
(e.g., Ferrarese \& Merritt 2000; Gebhardt et al. 2000a; Merritt \&
Ferrarese 2001a,b; Tremaine et al. 2002), even though those stars are
well beyond the radius of gravitational influence of the central black
hole.  This may imply a deep connection between the formation of galaxies
and the formation of supermassive black holes.

It has long been suspected that black holes of masses  $\sim
10^2-10^4\,M_\odot$ (intermediate-mass black holes, or IMBHs) may form
in, for example, the centers of dense stellar clusters (e.g., Wyller
1970; Bahcall \& Ostriker 1975; Frank \& Rees 1976; Lightman \& Shapiro
1977; Marchant \& Shapiro 1980; Quinlan \& Shapiro 1987; Portegies
Zwart et al. 1999; Ebisuzaki et al. 2001). However, for many years
there was no observational evidence for such a mass range.  In roughly
the last decade, X-ray and optical observations have revived this
possibility.  If such black holes exist, especially in dense stellar
clusters, they have a host of implications, particularly for cluster
dynamical evolution and the generation of gravitational waves.

Here we discuss the evidence for and implications of intermediate-mass black
holes.  There are two types of data that suggest the existence of
IMBHs.  First, there are numerous X-ray point sources, called ultraluminous
X-ray sources (ULXs), that are not associated with active galactic nuclei
yet have fluxes many times the angle-averaged flux of a $M<20\,M_\odot$
black hole accreting at the Eddington limit.  Second, several globular
clusters show clear evidence for an excess of dark mass in their cores.  At
present, we regard the X-ray evidence as more convincing, hence we discuss
ULXs in detail in \S~2. We discuss globular cluster observations in \S~3, as
well as models with and without IMBHs. In \S~4 we describe proposed
formation mechanisms for intermediate-mass black holes. In \S~5 we go
through models proposed for ULXs that do not involve IMBHs, and evaluate
several concerns that have been raised about the IMBH hypothesis. In \S~6 we
discuss the implications of IMBHs if they exist, with a focus on
gravitational radiation.  We conclude in \S~7 by listing a number of
important observations and theoretical calculations that will clarify our
understanding of these objects.

\section{Ultra-Luminous X-ray Sources}

\def\gapprox{\gta}
\def\lapprox{\lta}

Historically, most of the first detected bright X-ray sources were
identified as accreting neutron stars or black holes in our Galaxy
(Giacconi et al. 1971). However, it is not necessarily true that most
neutron stars and black holes are also strong X-ray sources.  Isolated
black holes emit a negligible amount of electromagnetic radiation, and
they are therefore very difficult to study.  If instead the black hole is
``active'' (i.e., accreting any significant amount of matter),  it is
usually an X-ray source, and the X-ray emission may be used to diagnose
the physical properties of the accretion. In X-ray sources with moderate
to high accretion rates, the accreting matter is believed to form a dense
accretion disk surrounding the black hole.  Geometrically thin disks are
usually thought to have an inner edge near the innermost stable circular
orbit, which is 3R$_s$ for a non-spinning (Schwarzschild) black hole.
Here, R$_s$ is the Schwarzschild radius, which is directly proportional
to the mass of the black hole $M$:
\begin{equation}
R_s = 2 {{G M}\over{c^2}} \approx 2.9 
\left({{M}\over{M_\odot}}\right)~{\rm km}.
\end{equation}
Note that magnetic links from the disk inside the innermost orbit could
allow energy extraction even inside the innermost stable circular orbit, as
might interaction with the spin of the black hole: for theoretical work see
Krolik (1999); Gammie (1999); Agol \& Krolik (2000); Reynolds \& Armitage
(2001), and for possible observational evidence see Wilms et al. (2001);
Miller et al. (2002b); Reynolds \& Nowak (2003). For a spinning
(Kerr) black hole, the last stable circular orbit, and thus the standard
inner disk radius, ranges from 0.5 to 4.5 R$_s$, depending on the black hole
spin and the sense of disk rotation (prograde or retrograde, e.g. Ori \&
Thorne 2000).

Gas in the inner disk interacts with itself, releasing energy and
transporting angular momentum, and much of the thermal
energy is emitted in X-rays (for stellar-mass black holes) and ultraviolet
light (for supermassive black holes in AGNs). Other physical processes in
coronal gas near the black hole also produce significant amounts of X-ray
emission, perhaps dominating the observed X-ray flux in AGNs.  Much of what
we know about black holes comes from  observations of ``active'' ones, and
therefore the study of black holes and the study of accretion-powered X-ray
sources are very closely linked.

\subsection{Isotropic Emission and the Eddington Luminosity}

Since one usually has no information about the flux radiation pattern
$f_X(\Omega)$ emitted by the X-ray source, it is common to assume
an ``isotropic'' X-ray
luminosity
L$_X$, as if the radiation pattern is uniform in all directions:
\begin{equation}
L_X = \int
\negthinspace
\negthinspace
\negthinspace
\negthinspace
\int d\Omega R^2 f_X(\Omega) = 4 \pi R^2 F_X,
\end{equation}
where F$_X$ is the observed X-ray flux, $R$ is the distance to the
source, and $f_X(\Omega)$ is the flux emitted in a particular direction.

The luminosity generated by accretion onto a black hole exerts an outward
radiation force on the accreting matter.  If the radiative acceleration
exceeds the acceleration of gravity, accretion is halted and no luminosity
is generated.  For accretion around a black hole, in which the matter is
highly ionized and electron scattering is the most important form of
opacity, a source of mass $M$ that accretes and radiates {\it
isotropically} therefore cannot have a luminosity that exceeds
the Eddington luminosity
\begin{equation}
L_E=
{{4\pi GMm_p}\over{\sigma_T}}=1.3\times 10^{38}\left({{M}\over{M_\odot}}
\right)\,{\rm erg\ s}^{-1}\; ,
\end{equation}
where $\sigma_T=6.65\times 10^{-25}$~cm$^2$ is the Thomson scattering
cross section.  We have assumed pure ionized hydrogen here; L$_E$ is 
slightly greater for a cosmic composition.  Therefore, if isotropy
holds, an observed flux places a lower limit on the mass of an
accreting black hole.

If instead the accretion or radiation are anisotropic, there is no
fundamental reason why the luminosity cannot exceed $L_E$ by an arbitrary
factor. Beaming of the radiation can produce a flux $f_X(\Omega)$ in a
particular direction that is much greater than the average over all
angles.  We will discuss  anisotropic models for ULXs further in \S~5, but
for now we concentrate on isotropic models.

\subsection{Definitions and Nomenclature}

If we consider stellar-mass BHs to have a maximum mass of
$\approx$20~M$_\odot$ (e.g. Fryer \& Kalogera 2001), then the  Eddington
luminosity of an  ``intermediate-mass'' BH is $\gapprox$ 3 $\times$
10$^{39}$ erg~s$^{-1}$. This limits the bolometric energy output of the
object.  The X-ray  luminosity in the 2$-$10 keV band, for example, will
be a factor of a  few$-$10 times smaller, and it will be dependent on
the metallicity as well. The lower limit to the X-ray luminosity for a
ULX is defined to be 10$^{39.0}$ erg~s$^{-1}$.  In practice, this limit
distinguishes the ``normal'' BH X-ray binaries (XRBs), with  
$L_X \lapprox$10$^{39.0}$
erg~s$^{-1}$ found in our Galaxy from the intriguingly more luminous
ULXs found in some nearby galaxies.  The upper  limit for L$_X$ for ULXs
is not specified, but  usually objects have L$_X <$ 10$^{40.5}$
erg~s$^{-1}$, and most of them have L$_X <$ 10$^{40.0}$ erg~s$^{-1}$.
Quasars, supernovae, and other galaxy nuclei are usually omitted,
although some workers (e.g. Roberts et al. 2001) include X-ray luminous
supernovae.

Ultra-Luminous X-ray sources (ULXs) were named as such by several Japanese
workers who analyzed spectra from the Japanese X-ray satellite ASCA
(Mizuno et al. 1999, Makishima et al. 2000, Kubota et al. 2002). Here,
``ultra-luminous'' is gauged with respect to ``normal'' X-ray  binaries.
The term ULX is now widely used for these intriguing off-nuclear sources.
Another term that is used is ``Intermediate-luminosity X-ray Objects,''
(IXOs), which simply indicates that their X-ray luminosities are
intermediate between those of ``normal''  stellar-mass BH~XRBs, and AGNs.

\subsection{A Historical Background: ULXs in the Einstein Era}

\begin{figure}[b!]
\vskip -0.23truein
\psfig{file=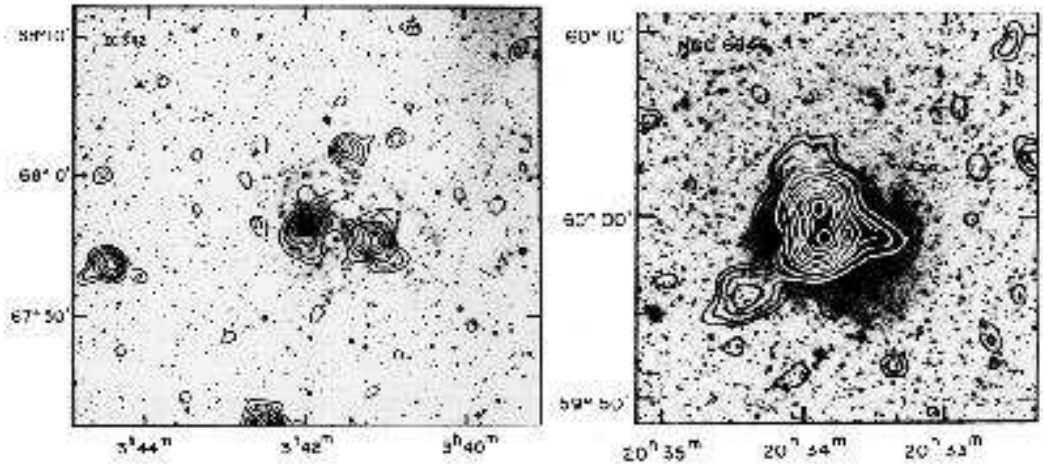,width=6.5truein,angle=270}
\vskip -0.23truein
\caption{Contours of the Einstein IPC X-ray emission from two nearby face-on
spiral galaxies (IC~342, on the left, and NGC~6946, on the right).  Both 
nuclei are quite luminous.  Even with only $\sim$1$^{\prime}$ resolution,
it is obvious that there are two very luminous, off-nuclear X-ray sources 
in IC~342 (west and north of the nucleus), and one in NGC~6946 (north of 
the nucleus).  Reprinted with permission from Fabbiano \& Trinchieri (1987).}
\label{colbertfig1}
\end{figure}

ULXs were observed as early as the 1980s, when  extensive X-ray
observations of external galaxies were first performed with the Einstein
satellite.  These studies  revolutionized the understanding of black holes
and their X-ray sources.  Many of the nearby AGNs in Seyfert galaxies were
expected to have very luminous X-ray nuclei, but it was a surprise to find
that many ``normal'' spiral galaxies also had central X-ray sources (see
Fabbiano 1989 for a review).   These X-ray sources had X-ray luminosities
$\gapprox$ 10$^{39}$ erg~s$^{-1}$, well above the Eddington value for a
single neutron star or a stellar-mass black hole.  The spatial resolution
of the most widely used instrument on Einstein (the Imaging Proportional
Counter, or the IPC, FWHM $\sim$1$^{\prime}$) is $\gapprox$1 kpc for
typical galaxy distances $\gapprox$4 Mpc, so it was not clear whether these
sources were single or multiple objects, or whether they were really
coincident with the nuclei.  We show some IPC images of the less ambiguous
cases in Figure~\ref{colbertfig1}.  Some  possibilities included a single
supermassive black hole with a low accretion rate, a black hole with a
normal accretion rate and super-stellar mass, hot gas from a nuclear
starburst, groups of $\gapprox$10 ``normal'' X-ray binaries (e.g. Fabbiano
\& Trinchieri 1987),  or very luminous X-ray supernovae (see Schlegel
1995). The supermassive black hole scenario was not very well supported
since there was not typically any other evidence for an AGN from
observations at optical and other wavelengths.  Another possibility was
that the errors in the galaxy distances were producing artificially large
X-ray luminosities. The nearest of these interesting X-ray objects is
located in the center of the Local Group spiral galaxy M33 (see Long et al.
1981). Several other Einstein observations of similar objects are reported
in Fabbiano \& Trinchieri (1987), and a summary of Einstein observations
are given in the review article by Fabbiano (1989).  Unfortunately, the
X-ray spectral and imaging capabilities of the IPC instrument were not
generally good enough to distinguish between the possible scenarios. Even
so, it was realized that these very luminous X-ray sources were
not uncommon in normal galaxies, and that they deserved further
attention.

\subsection{ROSAT observations of ULXs}

\begin{figure}[h!]
\psfig{file=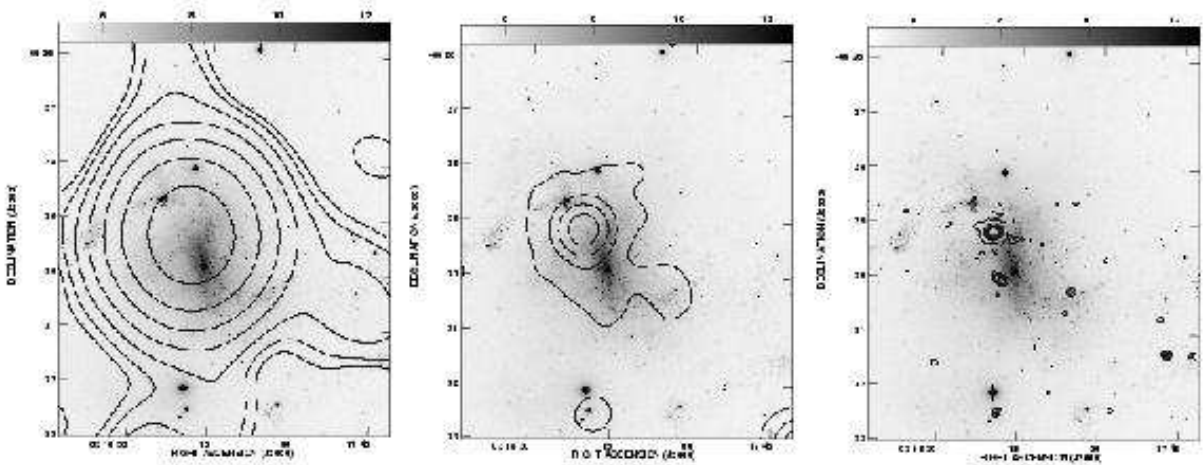,height=2.75truein,angle=270}
\caption{Comparisons between X-ray Instruments.  The greyscale image shows
optical I-band emission of the nearby face-on spiral galaxy NGC~1313 (from
Kuchinski et al. 2000).  The contours show X-ray emission near the center
of the galaxy from an Einstein IPC image (left), a ROSAT HRI image
(center), and a Chandra ACIS image (right).   Note the ambiguity of the
location of the ULX disappears as the resolution gets better, and many more
X-ray sources are found with the newer instruments on ROSAT and Chandra.
The I-band image was retrieved from NED, the Einstein and ROSAT images are
from NASA's HEASARC, and the ACIS image was kindly provided by G. Garmire.}
\label{colbertfig2}
\end{figure}

The ROSAT satellite was launched into orbit in  1990, and began
producing X-ray images at $\sim$10$-$20$^{\prime\prime}$ resolution. The
highest resolution instrument was the High Resolution Imager  (HRI; PSF
$\approx$ 10$^{\prime\prime}$). The sensitivity and spatial resolution
were a significant improvement over the Einstein IPC, many more ULXs were
discovered, and several surveys were done. It was soon found that some of
these luminous X-ray sources were not coincident with the galaxy nucleus.
For example, in Figure~\ref{colbertfig2}, we compare Einstein IPC and ROSAT HRI
(High Resolution Imager) images of the central X-ray
sources in the spiral galaxy NGC~1313. After registering the ROSAT image
with the X-ray bright supernova~1978K (X-3), Colbert et al. (1995) found
that the central Einstein source was actually located $\sim$1$^{\prime}$
($\sim$1 kpc) NE of the center of the nuclear bar. Some of the other
Einstein X-ray sources, are, however, still  consistent with being located
in the galaxy nucleus.  For example,  even with Chandra  accuracy
(1$^{\prime\prime}$), the M33 source is still coincident with the nucleus
of the galaxy, although it is not thought to be an AGN, since the dynamic
mass at that position is too small and the X-ray and optical properties
are more consistent with it being an XRB-like object (Gebhardt et al.
2001, Long et al. 2002, Dubus \& Rutledge 2002).

The PSPC spectrometer on ROSAT had much better spectral resolution than
the Einstein IPC, but it only covered soft X-ray energies (0.2$-$2.4 keV),
and so  was of limited use for diagnosing ULX emission models.  However,
much progress was made from surveys done with the ROSAT HRI. Four large
HRI surveys of nearby galaxies  (Colbert \& Mushotzky 1999, Roberts \&
Warwick 2000,  Lira, Lawrence \& Johnson 2000, and Colbert \& Ptak 2002)
showed that off-nuclear luminous X-ray sources were actually quite common
-- present in up to half of the galaxies sampled. At the time of the first
three surveys, ULXs were not a well defined class  of objects.  Roberts \&
Warwick (2000), Colbert \& Ptak (2002), Roberts et al. (2002a), and
ongoing work by Ptak \& Colbert indicate that ULXs, as we have defined
them in section 2.2, are present in one in every five
galaxies, on average.  When ROSAT survey work started showing that ULXs,
and thus possibly IMBHs, were quite  common, ULXs and IMBHs became a
popular topic of study.

\subsubsection{A Census of ULX Luminosities and BH Masses}

\begin{figure}[h!]
\hskip 1.5truein
\psfig{file=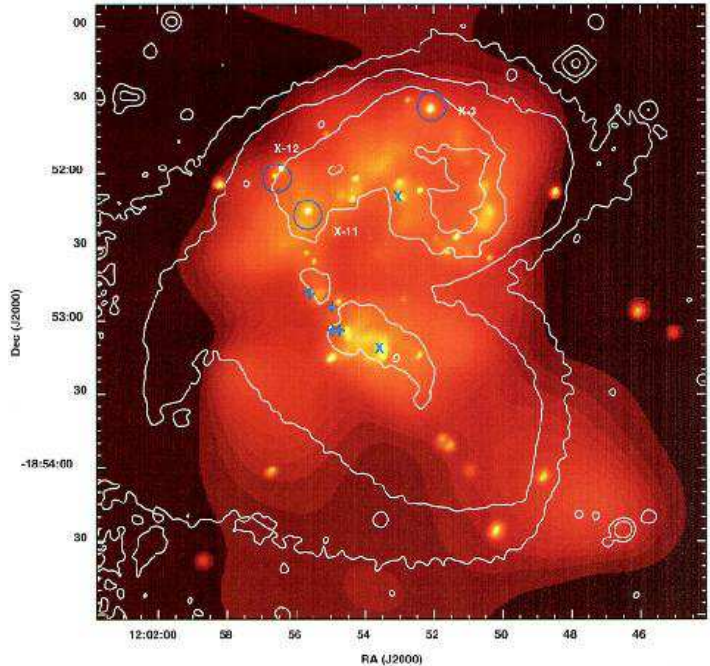,height=3.5truein,angle=270}
\caption{Adaptively smoothed Chandra ACIS image of the ``Antennae'' galaxies,
showing the ULXs and very luminous X-ray sources.  The white contours show
the optical emission levels of the galaxies.  The two nuclei NGC~4038 and
NGC~4030 are marked with crosses.  Reprinted with permission 
from Fabbiano et al. (2001). }
\label{colbertfig3}
\end{figure}

\begin{figure}[h!]
\psfig{file=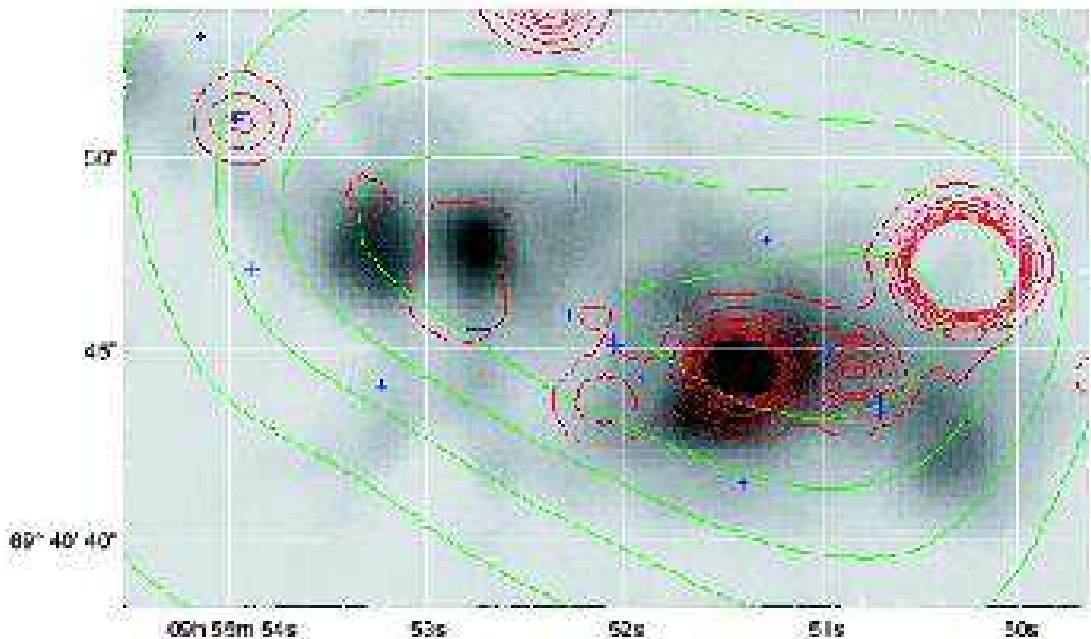,height=3.5truein,angle=270}
\caption{Gray-scale mid-infrared image of the 
central region of the edge-on irregular starburst
galaxy M82.  The dark contours show the X-ray
emission, in particular the famous ULX at the
far right of the image.  The light contours are
hard diffuse X-ray emission, and the crosses
are radio sources.  From Griffiths et al. (2000).}
\label{colbertfig4}
\end{figure}

As described above, IMBHs with M $\gapprox$20 M$_\odot$ have L$_E \gapprox$
3 $\times$ 10$^{39}$ erg~s$^{-1}$. In the Colbert \& Ptak (2002) catalog of
87 ULXs, 45 objects have 2$-$10 keV X-ray luminosities  L$_{X} >$ $3\times
10^{39}$~erg~s$^{-1}$. Eleven objects have L$_{X} >$
$10^{40}$~erg~s$^{-1}$, which corresponds to quasi-isotropic sources with
masses $M>70\,M_\odot$. Thus, the potential for IMBHs is clearly present.
Some galaxies, such as NGC~4038/9 (``The Antennae'') have $\gapprox$10 ULXs
with masses $\gapprox$10$-$(few)100 (by the Eddington argument), if the
X-rays are not beamed (see Figure~\ref{colbertfig3}, and Fabbiano et al.
2001, but see also Saviane, Hibbard, and Rich 2003 for a closer distance to
the Antennae, which implies lower luminosities). The brightest ULX yet
observed, in the galaxy M82 (see  Figure~\ref{colbertfig4}, and Ptak \&
Griffiths 1999; Matsushita et al. 2000; Kaaret et al. 2001), has a peak
X-ray luminosity of $9\times 10^{40}$~erg~s$^{-1}$ (Matsumoto et al. 2001),
implying a mass $M>700\,M_\odot$ by Eddington arguments.   This is well
beyond the expected $\sim 100-200\,M_\odot$ upper limit of stellar mass in
the current universe (see the introduction to \S~4).  In addition, even a
star that starts its life with a high mass may lose most of it to winds and
pulsations, leaving behind a black hole of mass $M\lta 20\,M_\odot$ if it
forms with roughly solar metallicity (e.g., Fryer \& Kalogera 2001).
Objects of such mass must either have accumulated most of their matter by
some form of accretion, or have formed in some other epoch of the universe.

The possibility remains that these objects are under-luminous supermassive
black holes.  However, as we explore in section 2.4.2,  the locations of
the ULXs within the galaxies rule against masses more than $\sim
10^6\,M_\odot$ in many cases.

\begin{figure}[h!]
\psfig{file=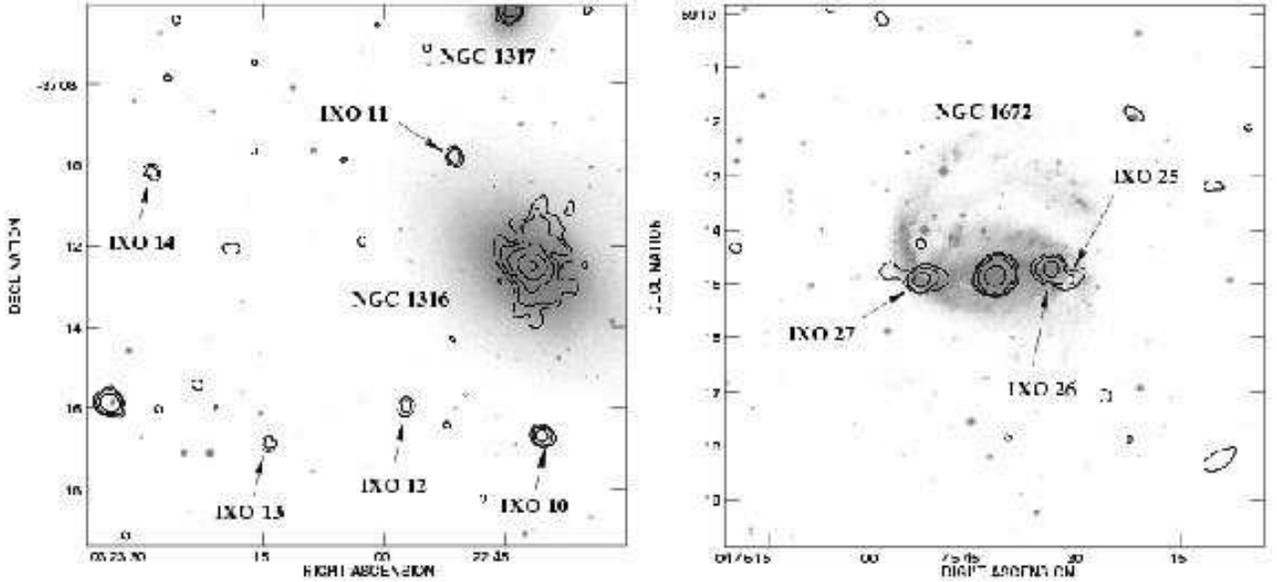,height=3.5truein,angle=270}
\caption{Two example galaxies with ULXs, from Colbert \& Ptak (2002).
On the left we show five of the six ROSAT HRI ULXs (IXOs) in the elliptical 
galaxy Fornax~A (NGC~1316).  On the right is the face-on barred spiral 
galaxy NGC~1672, which has two ULXs, both positioned at the end of the bar,
straddling the nuclear X-ray source.  }
\label{colbertfig5}
\end{figure}

Although the Colbert \& Ptak (2002) ROSAT HRI sample is the largest
published ULX catalog (see Figure~\ref{colbertfig5}), an even larger number 
of ULXs have
been found with Chandra.  Swartz, Ghosh, \& Tennant (2003),
Swartz et. al (2003, in prog.) and  Ptak \&
Colbert (2003, in prog.)  are finding $\sim$200$-$300 ULXs from analyses
of currently available  Chandra archive data.  This implies a factor of
$\sim$2$-$3 times more potential IMBHs than the ROSAT surveys estimate.

\subsubsection{Location in Galaxies}

\begin{figure}[h!]
\vskip -0.15truein
\hskip 1.5truein
\psfig{file=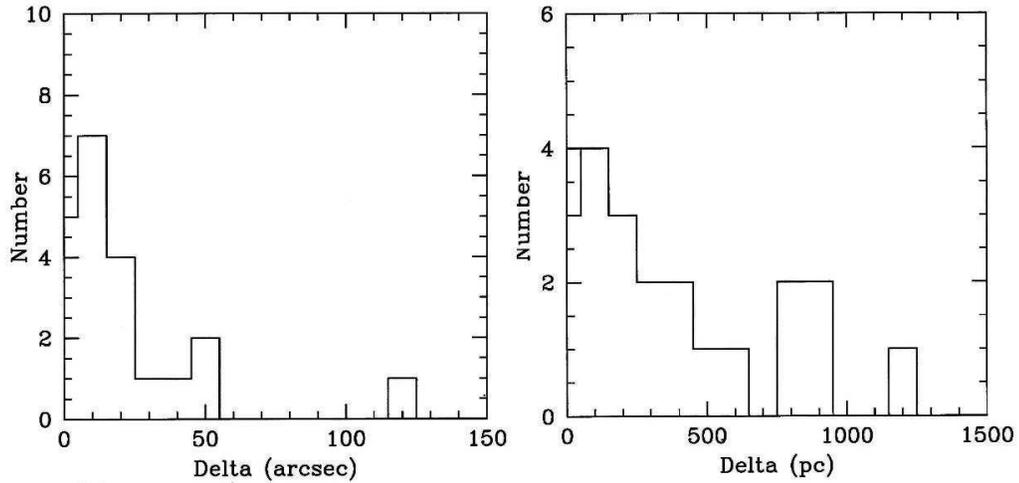,height=3.5truein,angle=270}
\vskip -0.23truein
\caption{Histograms of offset of near-nuclear ROSAT HRI X-ray sources from 
nucleus,
in arcseconds (left) and parsecs (right).  From Colbert \& Mushotzky (1999).
}
\label{colbertfig6}
\end{figure}

Colbert \& Ptak (2002) have compiled a list of all of the ULX candidates
observed with the ROSAT HRI.  This list shows that ULXs are found in both
spiral and elliptical galaxies, as well as a few irregular galaxies.  In
spirals,  Colbert \& Mushotzky (1999) find that the bright X-ray sources
are  often near, but clearly distinct from, the dynamical centers of the
galaxies, with an average projected separation of 390 pc 
(Figure~\ref{colbertfig6}). In
ellipticals, the ULXs are almost exclusively in the halos of galaxies, and
it is possible that these ULXs are distinct from those in spiral galaxies
(Irwin, Athey \& Bregman 2003). The off-center positions in spiral galaxies
show that these are not under-luminous supermassive black holes, because an
object of too large a mass would sink to the center via dynamical friction
in much less than a Hubble time.  More quantitatively, adopting equation
(7-27) of Binney \& Tremaine (1987), the dynamical friction time is

\begin{equation}
t_{\rm fric}\approx {5.0\times 10^9\,{\rm yr}\over{\ln\Lambda}}
\left({{r}\over{\rm kpc}}\right)^2
\left({{\sigma}\over{200\,{\rm km\ s}^{-1}}}\right)
\left({{M}\over{10^7\,M_\odot}}\right)^{-1}
\end{equation}
where $\sigma$ is the velocity dispersion, $r$ is the distance from the
dynamical center of the galaxy, and $\ln\Lambda\sim 5-20$ is the
Coulomb logarithm.  For example, for the most luminous X-ray source
in M82, $t_{\rm fric}\approx 10^{10}\,{\rm yr}(10^5\,M_\odot/M)$,
for an assumed $\sigma$ of 100 km~s$^{-1}$ (Kaaret et al. 2002).  
Thus, high masses are ruled out for ULXs that are near, but not located in,
the nucleus.  Technically, one must keep in mind
an individual object could have a much larger mass if it were
many kpc away from the galactic center, but it is usual and probably
correct to assume that this is quite a rare occurrence.

\subsection{X-ray Energy Spectra of ULXs}

As discussed previously, the earliest X-ray spectra of ULXs were obtained
with the Einstein IPC instrument (e.g., Fabbiano \& Trinchieri 1987).  These
spectral data were quite crude, however, and did not offer much in the way of
diagnostics of the nature of the X-ray emission.  Independent of any specific
physical models, state transitions have been observed from some ULXs (e.g.,
Kubota et al. 2001 observed transitions between soft and hard spectra for two
sources in IC~342).  These state transitions reinforce the identification of
ULXs with accreting black holes, although the transition behavior is
complicated (see the end of \S~2.7).  More detailed inferences, however, depend
on the spectral model used. A popular ULX model for ASCA spectra was the
multi-color disk (MCD) blackbody model, since it was commonly used to fit
X-ray spectra of ``normal'' BH XRBs.

\subsubsection{The Multi-color Disk Blackbody Model}

In the MCD model, each annulus of the accretion
disk is assumed to radiate as a blackbody with a radius-dependent
temperature, as modified by a spectral correction factor (Mitsuda et al. 1984).
The spectral flux $f(E)$ can be written as:
\begin{equation}
f(E) = {{\cos\theta}\over{R^2}} \int_{r_{in}}^{r_{out}} 2 \pi r B_E(T) dr,
\end{equation}
where 
$\theta$ is the angle of the disk axis with respect to the line of
sight, $R$ is the distance to the source, and $B_E(T)$ is the Planck function
at energy $E$.  Since $T(r) \propto r^{-3/4}$ for an assumed thin disk,
the flux can also be written in terms of $T$:
\begin{equation}
f(E) = 
{{ 8 \pi r_{in}^2 \cos\theta }\over{ 3 R^2 }} \int_{T_{in}}^{T_{out}} 
\left(
{{T}\over{ T_{in}}}\right)^{-{{11}\over{3}}} B_E(T) dT.
\end{equation}
In this model, the inferred temperature $T_{\rm in}$ of the innermost
portion of the disk is related to the mass of the black hole:
\begin{equation}
kT_{\rm in}\approx 1.2\,{\rm keV}\left(\xi\over 0.41\right)^{1/2}
\left(\kappa\over 1.7\right)\alpha^{-1/2}\left({\dot M}\over{{\dot M}_E}
\right)^{1/4}\left({{M}\over{10\,M_\odot}}
\right)^{-{{1}\over{4}}}
\end{equation}
(e.g., Makishima et al. 2000, eq. 10).  Here $\kappa\approx 1.7$ is
a spectral hardening factor, $\xi\approx 0.4$ is a factor that takes
into account that the maximum temperature occurs at a radius larger
than the radius of the innermost stable circular orbit, and $\alpha=
R_{\rm in}/(6GM/c^2)$ is unity for a Schwarzschild spacetime and
$\alpha=1/6$ for prograde orbits in a maximal Kerr spacetime.
Thus, if $T_{\rm in}$ inferred from MCD fits is representative, one expects
lower temperature from accreting IMBH than from accreting stellar-mass
black holes. 
Some detailed aspects of the application of the MCD model to ULX spectra
are given in Makishima et al. (2000).

\subsubsection{ASCA Spectral Modeling of ULXs and the 
``high temperature'' Problem}

Although ASCA had a poor PSF (FWHM $\sim$1$'$), it had far better
sensitivity and spectral resolution than the Einstein IPC, and had much
wider spectral coverage (0.4$-$10 keV) than the ROSAT PSPC.   Therefore,
substantial progress was made using ASCA observations of ULXs in nearby
galaxies. ULX ASCA spectra are often modeled with a (soft) MCD component
for the disk emission, plus a (hard) power-law component, which is
presumedly Comptonized disk emission (e.g. see Takano et al. 1994). As for
Galactic BH~XRBs, the power-law photon index $\Gamma$ was noticed to  be
hard ($\Gamma \approx$ 1.8) in ULX low-flux states, and soft  ($\Gamma
\approx$ 2.5) in ULX high-flux states (e.g.,  Colbert \& Mushotzky 1999,
Kubota et al. 2002).

The implications of the MCD model were, however, problematic. While X-ray
luminosities of $\sim$10$^{39-40}$ erg~s$^{-1}$ are simply explained
by an intermediate-mass black hole with sub-Eddington  accretion, the
temperature $kT_{in}$ (and radius $r_{in}$) of the inner accretion disk,
derived from MCD spectral fitting, are too  high (low) for IMBHs (Mizuno
et al. 1999; Colbert \& Mushotzky 1999; Makishima et al. 2000; Mizuno,
Kubota, \& Makishima 2001). BH XRBs with
stellar-mass black holes in our Galaxy typically have temperatures
$kT_{\rm in}\approx 0.4-1$~keV, while ULXs have  $kT_{in} \approx$
1.1$-$1.8 keV, which is more consistent with LMXB micro-quasars in our
Galaxy (e.g. Makishima et al. 2000). Given the large implied X-ray
luminosities of ULXs, one might expect them to have lower $kT_{\rm in}$
(Eq. 7).

One explanation of the high-temperature problem is to suppose that the
compact object is a stellar-mass BH with M $\lapprox$10 M$_\odot$, and
the X-ray emission is somehow beamed.  This may well be the right
model for some ULXs, but as we discuss below and in \S~5.1 there are 
individual sources with circumstantial evidence against beaming as well
as properties of ULXs as a class that are not yet fully addressed in
beaming scenarios.

Mizuno et al. (1999),  Makishima et al. (2000), and Ebisawa et al. (2001)
offer several potential explanations for the ``high temperature'' problem.
For example, it is possible that the BH is a Kerr IMBH, and the resulting
frame dragging can shrink the inner radius of the accretion disk up to
$\approx$6 times less than that of a Schwarzschild BH, for which r$_{in}
\gapprox$ 3 R$_{s}$.  Therefore, $r_{in}$ can be smaller, and $T_{in}$ is
larger, as implied by the MCD models.  Kerr models work  well for ULXs as
IMBHs (e.g., Mizuno et al. 2001), but imply very high disk inclination angles
(i $\gapprox$ 80$^{\circ}$), Ebisawa et al. 2001, Ebisawa et al. 2003).

One may also relax the assumptions of the
``thin disk'' model.  For example, increasing $\kappa$, the ratio of the 
color temperature to the effective temperature, will yield higher masses
(e.g. Shrader \& Titarchik 1999), and so will increasing the correction
factor $\xi$, which adjusts for the fact that $T_{in}$ occurs at a slightly
higher radius than $r_{in}$ (see Kubota et al. 1998).  The mass $M$ is 
proportional to the product $\kappa^2\xi$.  Makishima et al. (2000) shows
that $\kappa^2\xi$ has to differ largely from values for ``normal'' BH XRBs
for the ``high-temperature'' problem to be solved.

Finally, one may completely abandon the physically thin accretion disk
model. Abramowicz et al. (1988) and Watarai et al. (2000) show that very
high accretion rates  $\dot{M} \gapprox$ 10 L$_E/c^2$ lead to an ADAF
(Advection-Dominated Accretion Flow) solution (the so-called  ``slim
disk'' model), and that this can explain the  $r_{in} \propto T_{in}^{-1}$
relationship, found for MCD fits to ASCA spectra of ULXs (Mizuno et al.
2001).  The slim-disk model allows masses to be slightly larger
($\lapprox$10$-$30 M$_\odot$), but not as large as $\sim$100 M$_\odot$.

Much effort has gone into trying to explain why the MCD temperatures
$kT_{in}$ are so high for ULXs.  However, it is possible that the ULXs are
{\it not} well represented by a simple MCD disk model after all. For
example, when simulated spectra of accretion disks are fit with  MCD
models, $r_{in}$ and/or the disk accretion luminosity are very poorly
estimated (e.g. Merloni et al. 2000, Hubeny et al. 2001). In addition,
since the PSF of ASCA is so large, ASCA spectra can be contaminated by
diffuse  X-ray emission and by X-ray emission from other point sources
positioned  extraction regions, so that a single MCD model is
inappropriate.  In fact, as we now discuss, an increasing
number of ULX spectra  from the smaller PSF instruments on Chandra and XMM
now show much lower values of $kT_{in} \sim$ 0.1 keV.

\subsubsection{XMM and Chandra Modeling of ULX X-ray Spectra}

XMM and Chandra observations have the advantage that their spatial
resolution ($\approx$1$^{\prime\prime}$ for Chandra,
$\approx$4$^{\prime\prime}$ for the MOS2 camera on XMM) is good enough
that contamination from other X-ray sources is not as problematic as it
is for ASCA. They also have significantly better throughput than ASCA,
which improves the signal to noise, and the bandwidth is also greater,
which increases the flux from sources and allows detection of some ULXs
that are absorbed in soft X-rays.

While ASCA ULX spectral models often required both MCD and power-law
components, XMM and Chandra spectra are often fit with a single component
(either MCD or power-law).  This does not necessarily imply incompatibility
between the ASCA data and the XMM/Chandra data, because of the different
fields of view and sensitivities of the instruments.
Some anomalous ``super-soft'' ULXs emit
essentially all of their photons below a few keV, and do not always have
MCD spectra, or power-law spectra with slopes typical of XRBs. For example,
the spectrum of the super-soft ULX in NGC~4244 is  better fit with a  very
steep ($\Gamma \sim$ 5) {\it power-law} model (Cagnoni et al. 2003). It is
possible that these sources are quite different from most other ULXs.

In general, a simple power-law model with $\Gamma \approx$ 2 fits many XMM
and Chandra ULX spectra well. For example,  Roberts et al. (2001) show that
the ULX in NGC~5204 is best fit by a  power-law model.   Similar results
are found for the ULXs in NGC 3628 (Strickland et al. 2001) and the
Circinus galaxy (Smith \& Wilson 2001). Terashima \& Wilson (2003) observed
nine ULXs in M51 with Chandra and find that four are fit as well with a
power law as with an MCD, two are fit better with a power law, and
one is an emission line object in which a power law is assumed as
a continuum, compared to two super soft sources where an MCD fits
better than a power law. Roberts et
al. (2002) find that three of the five brightest ULXs in  NGC~4485/90 are
better or equally well fit by a power-law model, compared to the MCD
model.  Foschini et al. (2002a) examined eight ULX candidates with XMM and
find that the MCD model is never the best fit to the data; indeed, a power
law fits better in 5 of the 8 cases, although the statistics are poor.
Foschini et al. (2002a) also examined 10 other ULX candidates with data
too poor for spectral fits.  Two of the 18 total candidates have been
identified with background sources.  One of the power law sources
(NGC~4698 ULX 1) has been identified with a background BL~Lac object
at redshift $z=0.43$ (Foschini et al. 2002b), and one of the sources with
poor statistics (NGC~4168 ULX 1) has been identified with a background
starburst nucleus at $z=0.217$ (Masetti et al. 2003).  This injects a
cautionary note that some ULXs that are best fit with power laws may
actually be background nuclei.

There are also select cases for which a single MCD model is
preferred  over a power-law (e.g., M81 X-6 in Swartz et al. 2003, and also
some objects listed in references  above).  It is likely that both MCD and
power-law components are present, as in Galactic  BH~XRBs, but the quality
of the spectra are not good enough to statistically require the weaker
component (e.g. Humphrey et al. 2003).

\begin{figure}[h]
\hskip 1.0truein
\psfig{file=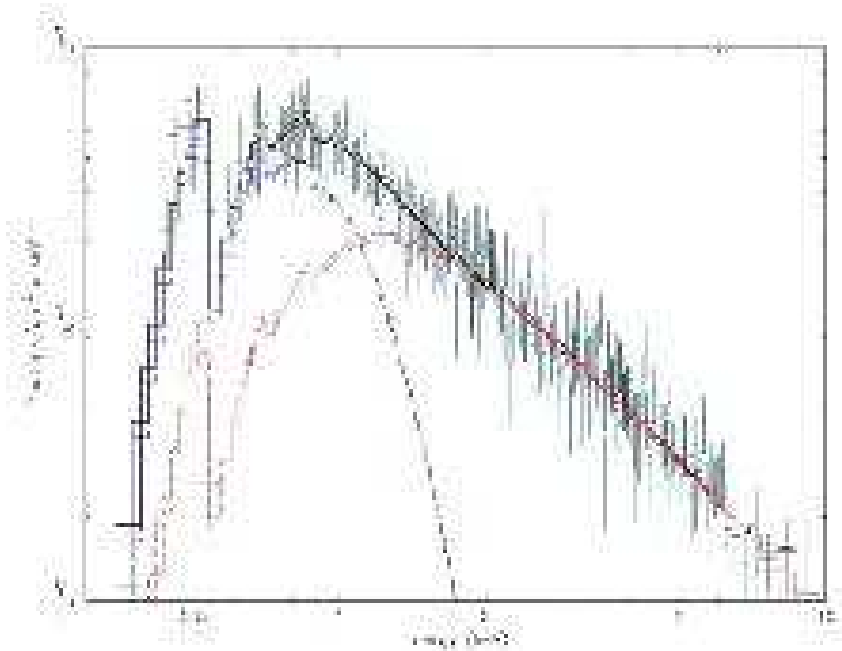,height=3.5truein,angle=270}
\caption{Unfolded XMM MOS spectrum of the ULX NGC~1313 X-1, along with 
components from an MCD model (blue) plus a power-law model (red).  Absorption
of soft X-rays below $\sim$1 keV is also modeled in the fit.  Reprinted
with permission
Miller et al. (2003a).}
\label{colbertfig7}
\end{figure}

An interesting result from the high-quality XMM ULX spectra is that, in
some cases if an MCD component exists, its inferred  temperature is much
less than the value obtained from ASCA observations  (some ROSAT
observations also suggested two-component fits, consistent with an IMBH;
see, e.g., Fabian \& Ward 1993). For example, Miller et al. (2003a)
analyzed XMM data of the brightest ULXs in NGC~1313, and found that a
two-component fit is necessary (see Figure~\ref{colbertfig7}), with an
inferred inner disk temperature $kT_{\rm in}=0.15$~keV.  In comparison,
Colbert \& Mushotzky (1999) analyzed two ASCA observations of NGC~1313 X-1;
one had a hard spectrum, with an MCD best fit temperature of  $kT=1.5$~keV,
while the other was softer and more consistent with the recent XMM
analysis. Some of the  ASCA spectra did actually imply low values for
$kT_{in}$.  For example, the ULX NGC~5408 X-1 is best fit with a MCD
temperature  kT$_{in} \approx$ 0.1 keV (Colbert \& Mushotzky 1999), and
this is confirmed with Chandra (Kaaret et al. 2003).  Similarly, the joint
ROSAT$+$ASCA fit of the X-ray spectrum of the ULX in Ho~II yields $kT_{in}
\approx$ 0.17 keV (Miyaji et al. 2001). XMM spectra of four of the
``Antennae'' ULXs are also consistent with cool MCD disks with kT$_{in}
\sim$ 0.1 keV, as are XMM spectra of M81 X-9 (Miller, Fabian, \& Miller
2003).  Di Stefano \& Kong (2003) also report a number of quasi-soft
sources with kT$<$300 eV that could be related to IMBHs.

It is important to recognize that these results do not {\it prove} that
the inner disk temperature is cool, but they do demonstrate that the
inference of high temperature from previous observations was unwarranted.
Continuum spectra can often be fit with a variety of models, with widely
different physical implications.  Therefore, the current data and fits are
consistent with the presence of intermediate-mass black holes, but do not
require their presence (although attempts to infer the mass from continuum
spectra are ongoing; see Shrader \& Titarchuk 2003).

While power-law spectra are often assumed to be associated with low/hard
states of BH~XRBs like Cyg~X-1, ASCA spectra of ULXs can also be fit with 
a strongly Comptonized disk model, associated with high/anomalous states in
some Galactic BH~XRBs (Kubota, Done, \& Makishima 2002).  As we discuss in
section 2.7, there is growing observational evidence from Chandra and XMM
observations that many ULXs may exhibit this anomalous high/hard behavior.

Although Fe~K lines (6.4$-$7.0 keV) are not usually strong in BH~XRBs,
Strohmayer \& Mushotzky (2003) found that the famous M82 ULX has a very 
broad Fe~K line in an XMM spectrum.  This is not easily explained by beaming
models and thus provides indirect evidence for an IMBH scenario.

We should note here that nearly all of the X-ray spectral modeling results
are derived for ULXs in spiral galaxies.  This is primarily due to the larger
distances of the nearby ellipticals (primarily in Virgo), and thus
the lack of available photons for spectral analysis.  If ULXs in ellipticals
are indeed a different class than those in spirals (e.g. King 2002), we 
might expect a difference in their X-ray spectral properties.  
Pioneering work by Irwin et al. (2003) suggests, in fact, that ULXs in
elliptical galaxies may have harder spectra than their counterparts
in spirals.

\subsection{ULXs and Host Galaxy Type}

Now that Chandra is in full operation, its combined imaging and spectral
capabilities have allowed the literature on ULXs to blossom.
The excellent imaging sensitivities of Chandra and XMM ensure
that one is likely to detect  an ULX in observations of nearby galaxies
$\gapprox$20\% of the time, for integrations of more than a few hours
(Ptak 2001). Even short ``snapshot'' observations with Chandra and  XMM
have detected a significant number of ULXs (Sipior 2003, Foschini et al.
2002a).

Chandra observations of the merger pair NGC~4038/9 (``the Antennae'')
revealed 8 ULXs (where the luminosities were estimated assuming a Hubble
constant H$_0 =$ 75 km~s$^{-1}$~Mpc$^{-1}$,  Fabbiano, Zezas \& Murray
2001).  Since the Antennae have very high  star-formation rates, this
seemed to suggest that ULXs are directly related to the young star
population.  In fact, many of the well-studied ULXs are  located in
starburst galaxies:  M82 (e.g. Kaaret et al. 2001 and references therein),
NGC~3628 (Dahlem et al. 1995, Strickland et al. 2001), and  NGC~253 (K.
Weaver, priv. comm.). Recently, a large number of potential ULXs have been
reported in the Cartwheel galaxy by Gao et al. (2003), although
variability studies are still needed to certify their ULX status.  In
addition, Zezas, Ward, and Murray (2003) report 18 ULXs in Arp 299.  It
was therefore conjectured that ULXs are a special type of  high-mass BH
XRB with beamed X-ray emission, and not IMBHs  (King et al. 2001).

\begin{figure}[h]
\hskip 1.5truein
\psfig{file=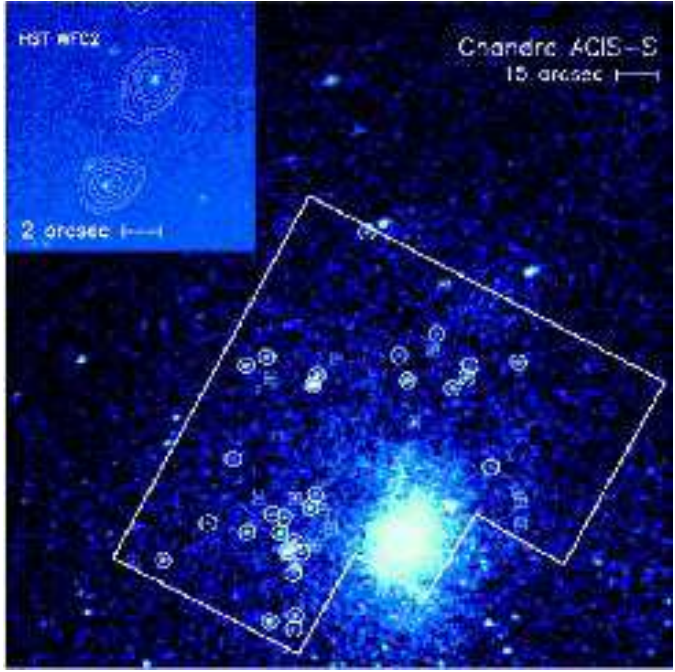,height=3.5truein}
\caption{Smoothed Chandra ACIS-S image of the
CD galaxy NGC~1399, with the
HST WFPC2 FOV overlaid.  The circles show the X-ray sources 
positions that are associated with globular clusters.  Reprinted with
permission from Angelini et al. (2001). }
\label{colbertfig8}
\end{figure}

However, further observations showed that not all ULXs were found in
starbursting, or even spiral galaxies.  For example, Angelini et al. (2001)
found several ULXs in a Chandra observation of the nearby giant elliptical
galaxy NGC~1399 (see Figure~\ref{colbertfig8}). Sarazin, Irwin, \& Bregman
(2001) find bright point sources up to $\approx 2.5\times
10^{39}$~erg~s$^{-1}$ in the elliptical galaxy NGC~4697, and conjecture
that although only 20\% of these sources are currently identified with
globular clusters, all the LMXBs may have originated in globulars.  In
general, low-mass X-ray binaries in early-type galaxies are strongly
correlated with globular clusters (Sarazin et al. 2003).  A census of ULXs
using all of the public ROSAT HRI data found that if one selects {\it only}
those galaxies with  detected ULXs, the elliptical galaxies with ULXs have
a larger number per galaxy than do the spiral galaxies  with ULXs (Colbert
\& Ptak 2002).   The elliptical galaxy NGC~720 has nine ULXs, which is
nearly as many that are found in the ``Antennae'' (Jeltema et al. 2003).
Since  ellipticals are also generally more massive than spirals, it does
not imply that they are more efficient at producing ULXs, but it does imply
that the  high-mass BH XRB scenario does not work for all ULXs, since
elliptical galaxies have virtually no young stars being formed.

As we discuss in the following sections, there are many possible
scenarios  that could explain  ULXs, and intermediate-mass black holes
remain an important contender. Chandra analyses of ULXs in elliptical
galaxies show that their X-ray spectra may be harder than those in spiral
galaxies  (Irwin, Athey \& Bregman 2003), corroborating that ULXs in
ellipticals are distinct from those in spirals. As the Chandra and XMM
data archives become more and more populated with spectral data for ULXs,
we will be able to better study their spectral and temporal properties,
and perhaps come to a better understanding of the underlying emission
processes, and the physical properties of their black  holes.

\subsection{X-ray Variability of ULXs}

It is possible that objects other than a single accreting black hole system
could produce X-ray luminosities $\ge$10$^{39}$ erg~s$^{-1}$.  For example,
some very young ($\lapprox$ 100 yr) supernovae are known to emit
$\sim$10$^{39}$ erg~s$^{-1}$ in X-rays.  However, their emission either
fades or remains constant on timescales of $\lapprox$1 yr (cf. Schlegel
1995). A cluster of $\sim$10 or more ``normal'' luminous XRBs could also
produce  $\sim$10$^{39}$ erg~s$^{-1}$.  However, neither of these scenarios
would account for the random or periodic variability that has been observed
in ULXs.

Colbert \& Ptak (2002) estimate random variability of  $\gapprox$50\% in
over half of all ULXs, eliminating supernovae or  XRB-clusters as likely
scenarios.  The brightest X-ray source in M82 brightened by a factor of 7
between two Chandra observations three months apart (Matsushita et al.
2000).  Long-term variability of ULXs on timescales of months to years
has been noted for ULXs in many nearby spiral galaxies: M81 (Ezoe et al.
2001, La~Parola et al. 2001, Wang 2002, Liu et al. 2002), Ho~II (Miyaji
et al. 2001), M82 (Ptak \& Griffiths 1999, Matsumoto \& Tsuru 1999, Kaaret
et al. 2001, Matsumoto et al. 2001), IC~342 (Sugiho et al. 2001, Kubota et
al. 2001),  Circinus (Bauer et al. 2001), NGC~4485/90 (Roberts et al.
2002a), M101 (Mukai et al. 2002), NGC 6503 (Lira et al. 2003), and
M51 (Terashima \& Wilson 2003).

Thus, variability on scales of months or longer is well-established.  For
periodic variability due to orbiting stars, Kepler's third law predicts very
short times for orbits near the BH:
\begin{equation}
P = 
2.00 \times 10^{-10} 
{ { ({a/{\rm km}})^{{3}\over{2}} }\over{ (M/M_\odot)^{{1}\over{2}} }} 
{\rm days}
=
3.65 \times 10^{2} 
{ { ({a/{\rm AU}})^{{3}\over{2}} }\over{ (M/M_\odot)^{{1}\over{2}} }} 
{\rm days}
\end{equation}
where $a$ is the semi-major axis of the stellar orbit.  With this in mind,
monthly X-ray monitoring can only sample orbits around $\sim$100 M$_\odot$
BHs for stars at radial orbits of $\gapprox$1 AU (1.5 $\times$ 10$^{8}$ km),
where the probability of eclipsing is quite low.  Thus, it is important to
test for periodicity on much shorter timescales, especially when searching
for evidence for IMBHs with M $\gapprox$ 100 M$_\odot$.

\begin{figure}[h!]
\hskip 1.5truein
\psfig{file=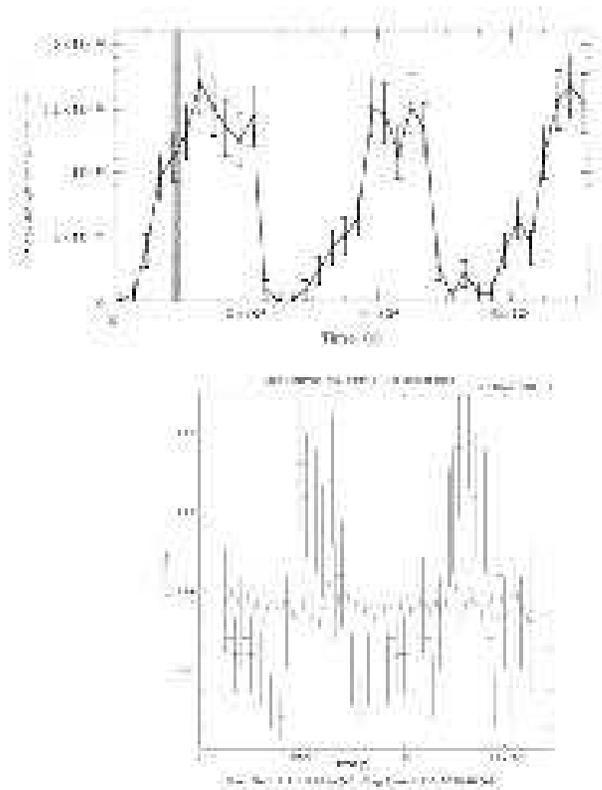,height=4.5truein}
\vskip -0.23truein
\caption{Chandra 
X-ray light curves of two ULXs that are candidates for periodic
behavior.  The upper figure shows
data for Circinus Galaxy X-1 (Reprinted with permission from Bauer et al. 2002).
Data for the shaded area have been interpolated.
The lower figure shows variation in M51 X-7, reprinted with permission
from Liu et al. (2002a).
The square data points in the center show the background level.}
\label{colbertfig9}
\end{figure}

There have been very few reports of variation on time scales less than a
few weeks.  There are currently three reported cases of variability on
time scales of hours, all of which have been interpreted as possibly
periodic (see Figure~\ref{colbertfig9}). Roberts \& Colbert (2003) report
aperiodic variability on timescales of a few hundred seconds from NGC 6946
X-11. Bauer et al. (2001) observed one source in the Circinus galaxy to
exhibit a count rate variation of a factor of 20, during a 67 ksec Chandra
observation. Three peaks are seen, which are consistent with a 7.5 hour
period. Bauer et al. (2001) discuss different mechanisms for this
variability, including eclipses, modulation of the accretion rate, or a
precessing jet.  Sugiho et al. (2001) observed a ULX in the spiral galaxy
IC 342 and found possible evidence for either a 31 hour or a 41 hour
period, admittedly based on only two peaks.  More recently, Liu et al.
(2002a) and Terashima \& Wilson (2003) report more than 50\% variation in
count rate from a ULX in M51, with a time of 7620$\pm$500 seconds between
the two peaks seen.

It is tempting to interpret these periods as orbital periods.
This would be highly constraining for the $\sim$2 hour period of the
M51 source, and would in fact imply that the companion is a 
$\sim 0.3\,M_\odot$ dwarf (Liu et al. 2002a).  However, it is
premature to draw conclusions because at this point no
source has been seen to undergo more than three cycles.  This is
a clear case in which sustained observations, especially of the
putative 2 hour period, are essential.  Only then will it be possible
to separate models in which the period is orbital (in which case it
should be highly coherent) from models in which the period is due to,
e.g., disk modes, in which the modulation could be quasi-periodic.

Variability on very short timescales (seconds to minutes) can be detected to
the same level of fractional rms amplitude as variability on longer
timescales, but the variability of sources from X-ray binaries to AGN tends
generally to decrease with increasing frequency.  This means that
variability at the few percent level would be detectable out to the Nyquist
frequency of observations of the brightest ULXs (e.g., to 1~Hz in the XMM
data analyzed for M82 by Strohmayer \& Mushotzky 2003).  It would be well
worth doing a systematic comparison of the broad-band power spectra of X-ray
binaries, ULXs, and AGN, given that one expects the maximum frequency at
which significant power exists to decrease with increasing mass.  Although
the lack of a fundamental theory of this variability limits our ability to
draw rigorous conclusions (e.g., the stellar-mass black hole LMC~X-3 has no
detected variation at $\nu>10^{-3}$~Hz; see Nowak et al. 2001), systematic
differences in the power spectra could provide insight into the nature of
ULXs.

The first, and so far only, quasi-periodic oscillation (QPO) in a ULX was
reported by Strohmayer \& Mushotzky (2003), based on XMM observations of
the brightest point source in M82.  They find a QPO at 54~mHz, with a
quality factor $Q\sim 5$ and a fractional rms amplitude of 8.5\%.  At the
time, the flux would imply a bolometric luminosity (if isotropic) of
$4-5\times 10^{40}$~erg~s$^{-1}$.  As discussed by Strohmayer \& Mushotzky
(2003), QPOs are usually thought to originate from disk emission,  which if
true makes this observation troublesome for a beaming interpretation.  This
is {\it not} because the frequency is low (for example, as mentioned by
Strohmayer \& Mushotzky 2003, a 67~mHz QPO has been observed with RXTE from
GRS~1915+105, which has a dynamically measured mass of $14\pm 4\,M_\odot$;
see Morgan, Remillard, \& Greiner 1997).  The problem is instead that if
the source is really a beamed stellar-mass black hole, the variability  in
the disk emission (which is nearly isotropic) would have to be of enormous
amplitude to account for the observations.  For example, even for a
$20\,M_\odot$ black hole accreting at the Eddington limit, the beaming at
$4-5\times 10^{40}$~erg~s$^{-1}$ would need to be a factor of $\sim 15$,
requiring intrinsic variability in the disk emission in excess of 100\%.
There are other sources in the XMM beam; the brightest of these sources has
an equivalent peak isotropic luminosity of $3.5\times
10^{39}$~erg~s$^{-1}$, comparable to the luminosity in the QPO of $3.4\times
10^{39}$~erg~s$^{-1}$  (Strohmayer \& Mushotzky 2003).  For this source to
produce the QPO  would therefore require nearly 100\% modulation, which
seems unlikely.  These observations therefore provide indirect evidence for
the IMBH scenario, although caution is still required because the theory of
black hole QPOs is not settled.  Recently, Cropper et al.
(2003) reported that the ULX NGC 4559 X-7 has a 28 mHz break in its
power density spectrum, which is consistent with a mass of a few
thousand solar masses, as is the measured thermal temperature of
kT=0.12 keV.

Combined spectral and temporal analyses can be a very powerful tool in
diagnosing ULX emission processes.  For example, ``normal'' BH XRBs often
exhibit a soft spectrum (with power-law slope $\Gamma \sim$ 2.5) in their
high state, and hard spectrum ($\Gamma \sim$ 1.8) in their low state. This
type of spectral variability has also been seen from ASCA observations of
several ULXs --- NGC~1313 X-1 (Colbert \& Mushotzky 1999), and two objects
in IC~342  (Kubota et al. 2001; see also Mizuno et al. (2001).  Some ULXs
observed with Chandra also show this behavior.  However, the opposite type
of  spectral variability (high/hard and low/soft) is also seen, such as for
NGC~5204 X-1 (Roberts et al. 2002a), and for four sources in ``the
Antennae'' (Fabbiano et al. 2003). Such ``anomalous'' spectral variability
has also been observed in some micro-quasars (e.g. GRS~1758-258, Miller et
al. 2002a).  Further spectral variability studies will certainly be useful
for understanding the ULX puzzle.

\subsection{Multiwavelength associations}

Since many starburst galaxies have ULXs, it is natural to search for clues to 
how ULXs are formed and fueled by studying their environment, and searching
for emission from possible companion star, accretion disk, and jet.  Thus,
observations of ULX fields at other wavelengths are very important.  ULX
environments could be young stellar clusters in starburst/spiral galaxies,
or globular clusters in spiral galaxy halos and in
elliptical galaxies.  

The superior spatial resolution ($\sim$1$^{\prime\prime}$) and absolute
astrometry ($\sim$1$^{\prime\prime}$) of Chandra has allowed matching of
the positions of X-ray sources with some optical sources. For spiral
galaxies, there may be an association of ULXs with star-forming regions
(e.g. Matsushita et al. 2000, Roberts et al. 2002b).  Young stellar
clusters associated with ULXs have masses  $\sim 10^4-10^5\,M_\odot$
(e.g., Zezas et al. 2002 for the Antennae galaxies; Matsushita et al. 2000
for M82).   However, ULXs are not always directly coincident with
star-forming regions  (Roberts et al. 2002a, Zezas et al. 2002). Since the
Antennae have so many ULXs, they can be used to determine exactly  how
frequently ULXs are associated with young star clusters. Zezas et al.
(2002) find eight ULXs possibly associated with 18 young stellar clusters,
where ``associated" means separated by less than 2". By randomly
scrambling the coordinates of the X-ray sources and clusters, Zezas et al.
(2002) estimate that by chance there would be $6\pm 2$ X-ray sources
associated with $8\pm 4$ optical sources, so the associations are still
tentative.  Intriguingly, Zezas et al. (2002) show that there is a small
but clear separation of typically 1-2" between a ULX and the nearest
young stellar cluster.  At the $\approx 20$~Mpc distance of the Antennae
(for $H_0=70$~km~s$^{-1}$~Mpc$^{-1}$), this corresponds to a physical
distance of 100-200~pc.  We will discuss in \S~4 and \S~5 how this
separation is interpreted in different models.

The first point-like optical counterpart to a ULX was found in NGC~5204 by
Roberts et al. (2001).  The optical source had a blue, featureless
spectrum and an optical luminosity typical of $\sim$8$-$20 O-giant or
supergiant stars. 
A young cluster of O
giants/supergiants supports the scenario of a high-mass BH XRB. Follow-up
HST imaging  work by Goad et al. (2002) showed that, in addition to the
luminous point source seen from the ground, there were two other fainter
point sources consistent with the X-ray source. Liu et al. (2002b) found
an optical counterpart to M81 X-11 with an optical luminosity and optical
color of an single O-star, suggesting that the  object is a high-mass BH
XRB.  Another point-like optical counterpart was found in HST images of a
ULX in the halo of the spiral  galaxy NGC~4565 (Wu et al. 2002).
However,  this optical counterpart appears to be a faint, blue globular
cluster, and is thus inconsistent with the high-mass BH XRB scenario.  HST
optical spectroscopic studies of the region surrounding the nearest ULX
(M33 X-8) were performed by Long et al. (2002).  They do not uniquely
identify the nature of the ULX, due to uncertainties in the X-ray
position.  This shows how complex optical follow-up work can be in crowded
regions of  spiral disk galaxies.  Very high precision X-ray astrometry
such as that of {\it Chandra} is needed to uniquely identify counterparts
in HST images, since even for short ``snapshot'' exposures in optical (B,
V, R, and I) bands, at least several optical sources are usually detected
within an  1$^{\prime\prime}$ radius circle in disk galaxies.

In elliptical galaxies, however, the astrometry problem is not as severe.
ULXs in elliptical galaxies are usually in the galaxy halo, which is sparsely
distributed with optical sources (globular clusters). Thus, identification of
a unique counterpart is often easier than in disk galaxies. For example,
Angelini, Loewenstein, \& Mushotzky (2001) performed a detailed comparison
between Chandra X-ray sources in the giant elliptical galaxy  NGC~1399 and
HST counterparts, finding that 26 of the 38 sources detected at $>$3$\sigma$
were obviously associated with globulars (Figure~\ref{colbertfig9}).   Two of
the three ULXs are associated with globulars. Other groups are also finding
that there is a strong correlation between X-ray sources in elliptical
galaxies and globular clusters (e.g. Kundu, Maccarone \& Zepf 2002). It will
be exciting to learn results from follow-up optical studies  to determine the
age, metallicity and other  derivable properties for these globulars, and of
other globulars with ULXs.

These results, combined with the results from starburst galaxies, show
that there is a strong link between ULXs and star clusters, whether they
be young star-forming regions, or globular clusters, which are 100$-$1000
times older. It is of interest that there are dozens of sources in globulars
around NGC~1399 with L$_X >$ 10$^{38}$~erg~s$^{-1}$, given that neither our
Galaxy (with $\approx$150 globulars, Harris 1996) nor M31 (with
$\approx$300$-$400 globulars, Hodge 1992, Fusi Pecci et al. 1993)  has any
X-ray sources in globular clusters with  L$_X \gta 10^{38}$~erg~s$^{-1}$
(Hut et al. 1993; Supper et al. 1997). Part of this may have to do with the
high number of globulars per unit mass around NGC~1399 (as it typical of
elliptical galaxies), which is 15 times the
average specific frequency for spiral galaxies such as the Milky Way and M31
(e.g., Kissler-Patig 1997), but there may also be evolutionary differences.

\begin{figure}[h!]
\psfig{file=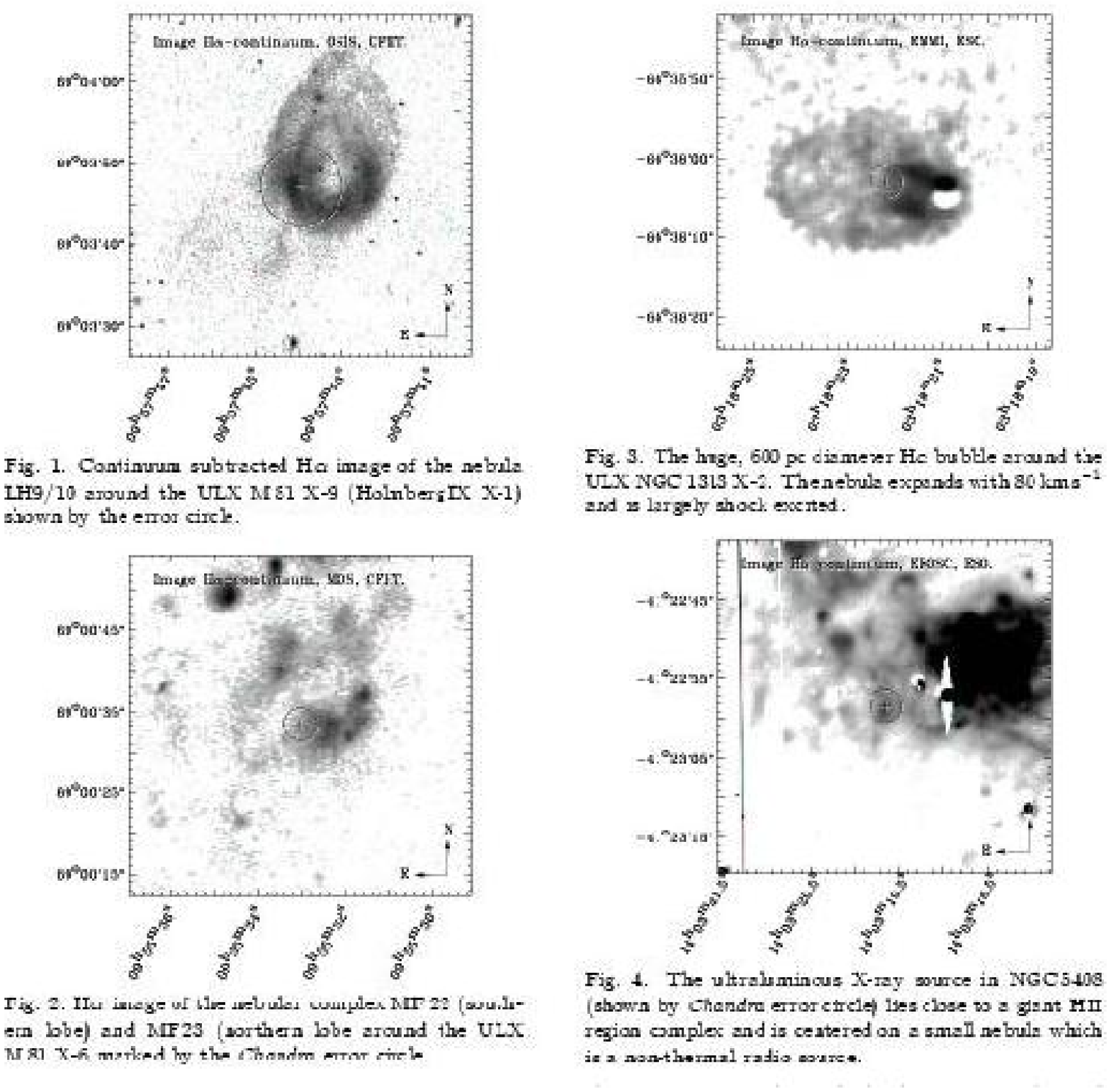,width=6.0truein}
\vbox{\vskip 0.2truein}
\caption{Examples of optical emission-line nebulae near ULXs.
In the top row are are continuum-subtracted
H$\alpha$ images of nebulae near Ho~IX X-1 and NGC~1313 X-2.
In the bottom row are H$\alpha$ images of nebulae near M81 X-6
and NGC~5408 X-1.  Reprinted with permission from Pakull \& Mirioni (2003).}
\label{colbertfig11}
\end{figure}

Studies of the environments around ULXs have led to some interesting
results. Some examples of ULX nebulae are shown in Figure~\ref{colbertfig11}. 
Pakull \&
Mirioni (2002) find that the ULX in the dwarf galaxy Holmberg II has an
optical nebula around it with substantial He II 4686\AA\ emission.  This
line is produced by the recombination of fully ionized helium, which
requires for its excitation a high-energy source.  Since the optical
luminosity is isotropic, it  places a lower limit on the X-ray luminosity
illuminating it; if the solid angle subtended by the nebula as seen from
the X-ray source is less than $4\pi$ then the true X-ray luminosity is
greater than inferred from the optical light, but it cannot be
significantly less.  Based on models of X-ray reprocessing where the X-ray
source is located inside the nebula, Pakull \& Mirioni (2002) conclude that
the optical radiation is consistent with an isotropic X-ray source and not
with significant beaming.  However, there is substantial uncertainty in the
correction factor from optical line flux to X-ray luminosity, so work of
this type needs to be repeated for a number of sources in order to draw
firmer conclusions.  Integral field spectroscopic observations by Roberts
et al. (2002a) indicate that many of the ULXs are actually located in
cavities free from optical line-emitting gas, although it is not clear
whether this is due to the absence of gas (e.g., gas cleared away by
shocks), or to highly ionized gas irradiated by the ULX.  Optical spectral
analyses of some of these ULX nebulae show evidence for both shocks and
photo-ionization (Pakull \& Mirioni 2003). This is intriguing, as there are
now at least two ULXs that are highly  variable (and thus are accreting
compact objects), but are directly associated with optical supernova
remnants (IC~342 X-1, Roberts et al. 2003, and MF16 in NGC~6946, Roberts \&
Colbert 2003; note that the precise mechanism for the ionization is not
rigorously established in these cases, and that jet ionization is a
possible alternative to supernova shock ionization).  Future
multiwavelength studies of optical ULX nebulae and ULXs in SNRs may provide
important clues as to how  ULXs form, or at least how they become
``active'' X-ray sources.

Radio counterparts to ULXs are only just starting to be found. The first
identification was reported for  NGC~5408 X-1 by Kaaret et al. (2003), with
an inferred 5~GHz radio power of $\sim$10$^{14}$ W~Hz$^{-1}$, consistent with
relativistically beamed jet  emission. If these ULXs are analogous to
Galactic ``micro-quasars'' (XRBs with relativistic jets), which
coincidentally have transverse jets, they could represent a class of
``micro-blazars'' that have their jets oriented directly toward the
observer.  Follow-up surveys are now underway to determine if
relativistically beamed radio jets are common in ULXs.

\section{Black Holes in Globular Clusters and as MACHOs}

\subsection{Kinematics of globular clusters}

As indicated in the previous section, although X-ray observations suggest
the existence of black holes of masses $M\sim 10^2-10^4\,M_\odot$ there is
as yet no direct measurement of the masses.  More direct measurements
might be obtained by optical observations of globular clusters.  It has
long been speculated (e.g., Frank \& Rees 1976) that the centers of
globulars may harbor $\sim 10^3\,M_\odot$ black holes.  If so, the massive
black holes affect the distribution function of the stars, producing
velocity and density cusps.  Unfortunately, observation of these cusps is
difficult.  For a central number density of
$10^5-10^6$~pc$^{-3}$, typical of dense globulars (Pryor \& Meylan 1993),
the projected surface density at 10~kpc is $\sim 10^2-10^3$ per square
arcsecond, more than can be resolved easily with even the {\it Hubble}
Space Telescope. Moreover, unlike the bright stars at our Galactic Center, which
have been observed for a decade to provide superbly precise measurements
of the central black hole mass, the stars in globulars are old and dim.

In addition, the radius of influence of an intermediate-mass black hole is
much smaller than it is for a supermassive black hole. For example, the
velocity dispersion near the center of our Galaxy is $\sim
100$~km~s$^{-1}$ (e.g., Tremaine et al. 2002), compared with a typical
velocity dispersion of $\sim 10$~km~s$^{-1}$ for a globular cluster (e.g.,
Pryor \& Meylan 1993).  The radius at which the orbital velocity around a
black hole of mass $M$ equals a velocity dispersion $\sigma$ scales as
$M/\sigma^2$, hence the radius of influence of the $\sim 3\times
10^6\,M_\odot$ black hole at the Galactic center is 30 times the radius of
influence of a $\sim 10^3\,M_\odot$ black hole in a globular cluster.  At
a distance of 10~kpc, a $10^3\,M_\odot$ black hole would influence
orbits within $\approx$1", making observations very challenging but not
impossible.  Therefore, ground-based adaptive optics observations, along
with space-based observations, have been applied to globulars to search
for massive black holes.

The current evidence is promising but not yet compelling; based on {\it
Hubble} observations, Gebhardt, Rich, \& Ho (2002) report a stellar
distribution in the M31 globular cluster G1 that favors the presence of a
$\approx 2\times 10^4\,M_\odot$ black hole at the 1.5$\sigma$ significance
level, and van der Marel et al. (2002) and Gerssen et al. (2002) find
marginal (0.7$\sigma$) evidence for a $\sim 2-3\times 10^3\,M_\odot$ black
hole in the center of the Galactic globular M15.  Given that these results
require extensive and careful modeling of the stellar distributions in
order to quote a significance, it is not yet possible to claim evidence for
intermediate-mass black holes in these systems.  Furthermore, recent n-body
modeling has shown that the evidence for a black hole may be even weaker
than thought previously for both M15 (Baumgardt et al. 2003a) and G1
(Baumgardt et al. 2003b).  Baumgardt et al. agree that there is an excess
of dark mass in the centers of these clusters, but suggest that it may be
modeled well by a cluster of lower-mass objects such as white dwarfs,
neutron stars, or stellar-mass black holes.

\begin{figure}[b!]
\hskip 0.3truein
\psfig{file=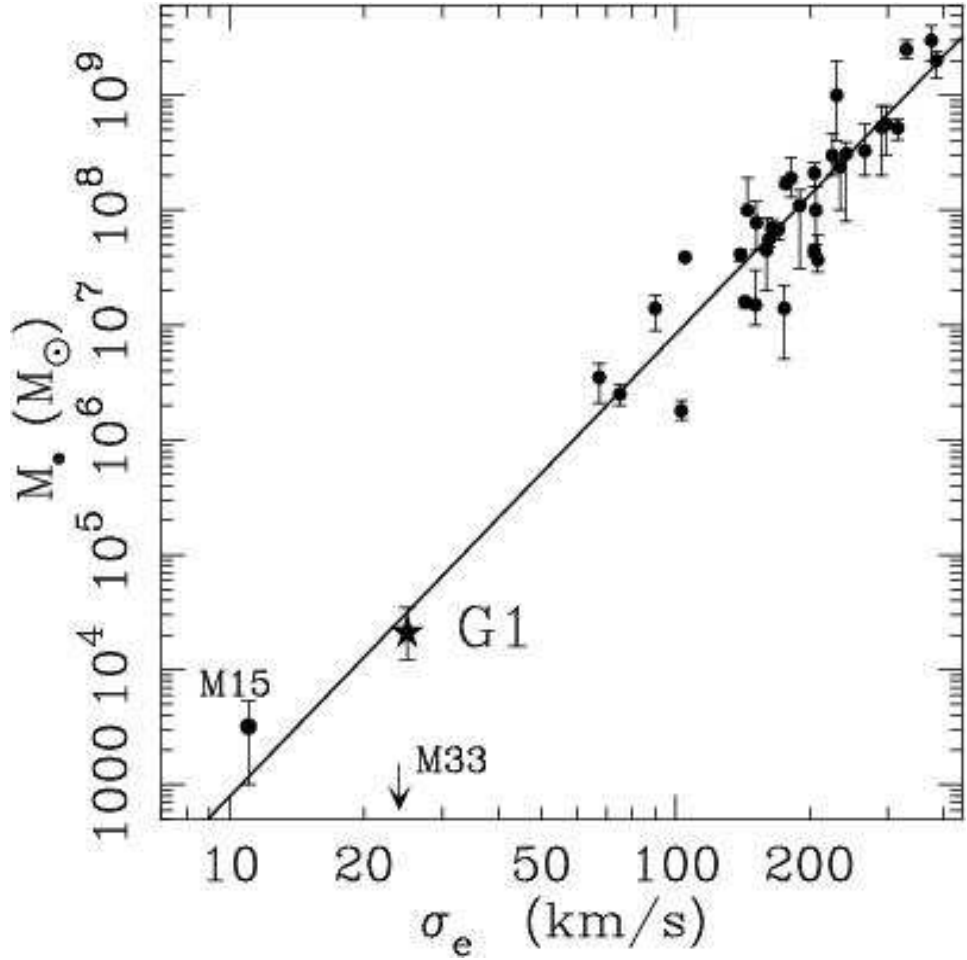,height=5.0truein,width=5.0truein,angle=270}
\caption{Relation between central velocity dispersion and estimated
black hole masses, for a number of galaxies as well as two candidate
IMBHs in globular clusters.  Reprinted with permission from Gebhardt 
et al. (2002).}
\label{millerfig1}
\end{figure}

As pointed out by Gebhardt et al. (2002), van der Marel et al. (2002),
and Gerssen et al. (2002), it is tantalizing that, taken at face value,
the best fit masses for black holes in G1 and M15 fall directly on the
extension of the relation
\begin{equation}
\begin{array}{rl}
M&\approx 10^8\,M_\odot (\sigma/200\,{\rm km\ s}^{-1})^4\\
&\approx 800\,M_\odot (\sigma/10\,{\rm km\ s}^{-1})^4\\
\end{array}
\end{equation}
between black hole mass $M$ and velocity dispersion $\sigma$ found for
supermassive black holes in galaxies (Ferrarese \& Merritt 2000; Gebhardt
et al. 2000; Merritt \& Ferrarese 2001a,b; Tremaine et al. 2002; see
Figure~\ref{millerfig1} for the $M-\sigma$ relation including two possible
IMBHs in globular clusters, from Gebhardt et al. 2002). Further
examination of density and velocity distributions in globulars is
obviously of the  highest importance in understanding intermediate-mass
black holes and possibly the formation of supermassive black holes.

\subsection{Modeling of millisecond pulsar distributions}

One promising path to dynamical detection of black holes in globulars
involves detailed modeling of the properties of individual objects. For
example, D'Amico et al. (2002) observe five millisecond pulsars in the
Galactic globular NGC~6752.  Three of these are in the cluster core: two of
these three have negative period derivatives and one
has an anomalously high positive period derivative.  From this, Ferraro et
al. (2003) conclude that the mass to light ratio in the core is likely to be
$M/L\approx 6-7$, much higher than inferred for most globulars.  If the spin
derivatives are ascribed to the overall gravitational potential of the
cluster, Ferraro et al. (2003) find that this implies the presence of
$1000-2000\,M_\odot$ of underluminous matter within the inner 0.08~pc of the
cluster.  Possibilities for this matter include an exceptional concentration
of dark remnants, a $\sim 1000\,M_\odot$ black hole in the center of the
cluster, or a $\sim 100\,M_\odot$ black hole that is offset but near the
projected location of the three millisecond pulsars.  The high spatial
resolution in the Ferraro et al. (2003) observations demonstrates that there
is no cusp down to 0.08~pc, implying that any central black hole has to have
a mass $M\lta 1000\,M_\odot$ and be within 0.08~pc of the core. 

Colpi, Possenti, \& Gualandris (2002) and Colpi, Mapelli, \& Possenti (2003)
focus on another of the five pulsars, which is in a binary with a
$0.2\,M_\odot$ star and is a remarkable 3.3 half mass radii away from the
cluster center. After surveying various mechanisms to produce this enormous
offset, they conclude that the ejection could be produced either by a
black hole binary in which the black holes are large but still in the
stellar range, or by an intermediate-mass black hole with a lighter
companion such as a stellar-mass black hole or a collapsed star.  Further
monitoring will likely be necessary to understand this system better.
In particular, study of the nature of the companion of the offset pulsar
will help, as will additional timing studies of the core millisecond
pulsars to detect or constrain the existence of wide binary companions.

\subsection{Core rotation in globular clusters}

\begin{figure}[b!]
\vbox{
{
\psfig{file=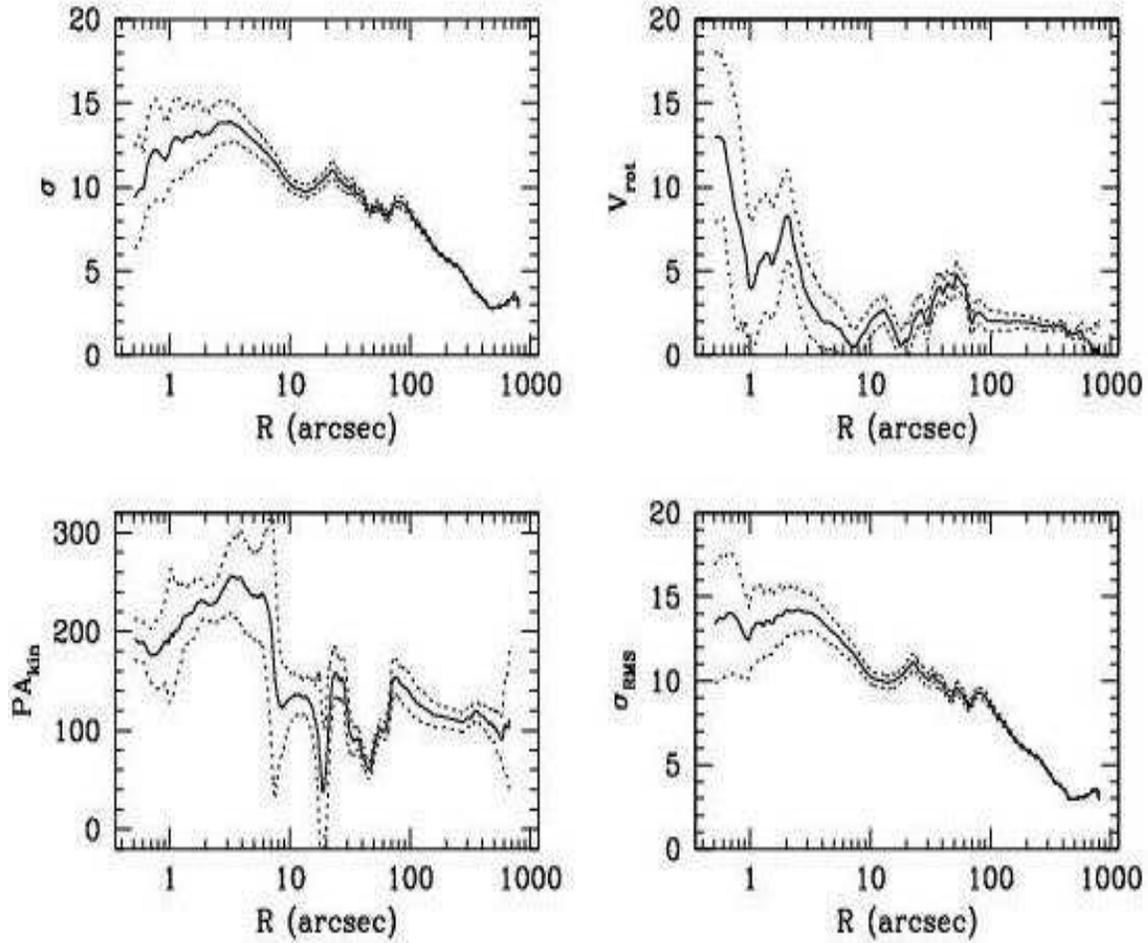,height=5.0truein,width=6.0truein,angle=90}}}
\caption{Velocity structure in the core of the Galactic globular
cluster M15.  Top left: random component of the line of sight
velocity as a function of angular distance from the center of the
cluster.  Top right: systematic rotational velocity as a function of angular
distance from the center.  Note that $v_{\rm rot}/\sigma\approx 1$ near
the core, indicative of significant net rotation.  Bottom left: position
angle of the rotation, as a function of angular distance from the center.
The position angle varies significantly, suggesting that the cluster as
a whole does not have a constant direction of rotation.  Bottom right:
total velocity dispersion, including both random and rotational components.
Reprinted with permission from Gerssen et al. (2002).}
\label{millerfig2}
\end{figure}

Another potentially exciting observational development in the dynamics  of
globulars has to do with rotation in the core, for which a possible
explanation involves an IMBH in a binary system with a stellar-mass black
hole.  N-body models of globulars without massive compact objects predict
essentially no rotation in the cores of clusters (see, e.g., Figure~3 of
Baumgardt et al. 2003b, which shows the expected rapid decrease of
ellipticity inside of $\sim$1~pc). However, Gebhardt et al. (2000b) and
Gerssen et al. (2002) find that in the center of M15 the rotational speed
is comparable to the velocity dispersion, $v_{\rm rot}/\sigma\approx 1$
(see Figure~\ref{millerfig2}, from Gerssen et al. 2002),
with a correspondingly high ellipticity of isophotes (K. Gebhardt, personal
communication). This might be produced if the entire cluster has a net
rotation, but in M15 the position angle of the rotation in the core is
different by 100$^\circ$ from the position angle further out, and indeed
the derived position angle wanders significantly with increasing distance
from the center.

As discussed in, e.g., Gebhardt et al. (2000b), typical n-body results
(e.g., Einsel \& Spurzem 1999) show that after an initial
increase in the central rotation velocity during collapse (likely driven
by the gravo-gyro instability; Hachisu 1979) the rotation velocity stabilizes,
and at no point does $v_{\rm rot}/\sigma$ approach unity.  If net rotation
is introduced artificially into the core, the angular momentum is 
transported outwards on a timescale comparable to the core relaxation
timescale, which can be $10^{7-8}$~yr for dense clusters.  The central
rotation may be increased in multimass models with net cluster rotation
(Arabadjis \& Richstone 1998), but if there is net rotation it is difficult
to understand the substantial variation in position angle far from the center.

An alternative that is consistent with the data is the presence of a
massive black hole binary in the core (F. Rasio, personal communication).
In M15, the evidence for rotation is based on $\sim 10$ stars in the
central 0".5 (Gerssen et al. 2002) with a probable total mass of
$\sim 5\,M_\odot$.  At the $\sim$10~kpc distance of
M15, 0".5 is 0.025 pc.  Suppose that the central mass is $10^3\,M_\odot$.
Then the total angular momentum in the central stellar system is
$L\approx 5\,M_\odot\sqrt{G(10^3\,M_\odot)(0.025~{\rm pc})}$.  This is
the amount of angular momentum that must be deposited in the center
(unless the rotation is shared by a large mass in unseen stars).
A $\sim 20\,M_\odot$ black hole orbiting the $10^3\,M_\odot$ central mass at a
semimajor axis $a\sim 10^{-3}$~pc has the required angular momentum.
Therefore, a $10^3\,M_\odot - 20\,M_\odot$ black hole binary has enough
angular momentum to account for the observed rotation.  As such a binary
hardens through three-body interactions, its orbital plane will change,
but the net result after it becomes a tight binary is that the original
angular momentum must have been transferred to the surrounding stars.
Eventually, the binary will merge, then another binary will form and
give its angular momentum to the stars.  Successive binaries have no
reason to have the same orbital planes, therefore one would expect that
the position angle will vary randomly.  

Note that although the original binary is much smaller than the size of
the rotating region, the binary will wander by $\sim 0.02-0.03$~pc for
$M\approx 10^3\,M_\odot$ (Merritt 2001; see below), hence as the binary
wanders it can deliver its angular momentum to the observed region.  As a
further consistency check, we may ask whether $10^{-3}$~pc, or roughly
200~AU, is a reasonable size for the initial binary.  If initially the
massive black hole is solitary, it can capture companions by three-body
exchange.  If the original binary (containing a $20\,M_\odot$ black hole
and a smaller companion) is hard, it has an orbital radius less than  $200
(\sigma/10~{\rm km~s}^{-1})^{-2}$~AU.  When captured by the black hole,
the binary will separate if it is larger than its Hill sphere at closest
approach (i.e., if the tidal acceleration across the binary exceeds the
gravitational acceleration of the binary itself).  For a closest approach
$r_p$, the critical radius is $r_H\approx (m/3M)^{1/3}r_p$, where $m$ is
the total mass of the binary and $M$ is the mass of the large single black
hole.  For $m\sim 20\,M_\odot$ and $M\sim 10^3\,M_\odot$, this implies
$r_p\sim 10a$, where $a$ is the initial semimajor axis of the binary.
Therefore, the $10^3\,M_\odot - 20\,M_\odot$ black hole binary can have an
initial orbital radius up to $\sim 10^{-2}$~pc, which provides plenty of
angular momentum.

Analysis of other clusters is underway, with some preliminary indications
that many have strong signatures of core rotation (K. Gebhardt, personal
communication).  If so, this is an exciting development.  Not only would
this provide good evidence of intermediate-mass black holes in globulars,
but the rapidity with which angular momentum is transported out of the
cores (in a core relaxation time, typically $10^{7-9}$~yr) means that for
us to see rotation now, there must have been a recent binary hardening
event.  This in turn implies that there are frequent mergers, perhaps
tens to hundreds per globular in a Hubble time, and therefore these may
be outstanding sources of gravitational radiation (see \S~6).  These data
must thus be examined with special care, to make sure that there are
no misleading systematics that give incorrect signatures of rotation.
In addition, focused n-body modeling will be crucial to see whether there
is any other way, without net rotation of the cluster, to get selective
rotation of the core with a varying position angle.

\subsection{Possible X-ray observations of IMBHs in globulars}

If a $\sim 10^2-10^4\,M_\odot$ black hole exists in a globular, there
are other possibilities for detection.  Although the original component
of gas in globulars is thought to have been evacuated, winds from
the $\sim 2-3$\% of stars currently on the red giant branch produce a
tenuous interstellar medium.  Bondi-Hoyle accretion onto the central black
hole can then produce visible emission in various bands, the most prominent
perhaps being X-rays and radio.  This accretion will happen far below the
Eddington rate, and in such a regime it is difficult to predict the
overall efficiency with which luminosity is generated, due to uncertainties
about the correct model of accretion in the low accretion rate limit
(e.g., advection-dominated accretion flows: Ichimaru 1977; Rees et al.
1982; Narayan \& Yi 1994; wind solutions: Blandford \& Begelman 1999;
Quataert \& Gruzinov 2000).
Nonetheless, one expects that a central black hole in a globular will be
a faint and otherwise unidentified source.  Depending on the mass of the
black hole, the source may or may not be at the dynamical center of the
globular.  A black hole in the mass range of interest wanders around the
core because of interactions with individual stars.  From equation
(90) of Merritt (2001), the expected wander radius of a binary black hole
of total mass $M$ in a globular cluster with a core radius $r_c$ and field 
stars of mass $m_f$ is
\begin{equation}
r_w\approx 0.22\,{\rm pc}(20 m_f/M)^{1/2}(r_c/1\,{\rm pc})\; .
\end{equation}
Grindlay et al. (2001) observed the Galactic globular 47 Tuc with
{\it Chandra} and found one unidentified X-ray source within the
wander radius of a $\sim 500\,M_\odot$ black hole ($\sim$2" at the
4~kpc distance of 47 Tuc) that had the right X-ray flux for Bondi-Hoyle
accretion with an efficiency of $10^{-4}$, characteristic of low radiative
efficiency flows.  This is intriguing, but more than one example will
be needed before conclusions can be drawn.  For example, Ho, Terashima,
\& Okajima (2003) find only an upper limit to any X-ray emission from
the dynamical center of M15.

\subsection{Detection of IMBHs from gravitational microlensing}

A final semi-direct method of detecting the masses of intermediate-mass
black holes is through gravitational microlensing.  Collaborations such
as MACHO (Alcock et al. 2001a), OGLE (Udalski, Kubiak, \& Szymanski
1997),  and EROS (Afonso et al. 2003) have been monitoring the Galactic
bulge and the Magellanic clouds for more than a decade, looking for the
achromatic signatures of microlensing.  For a fixed speed of a lens, more
massive objects produce longer events because their Einstein radii are
larger.  The longest such events introduce a parallax signal due to the
orbit of the Earth, which allows some breaking of degeneracies in the
parameters of the lens.  

Bennett et al. (2002a) report evidence of six events with estimated lens
masses greater than $1\,M_\odot$, with the most massive being
$6^{+10}_{-3}\,M_\odot$ and $6^{+7}_{-3}\,M_\odot$. Another event,
detected by both the MACHO and OGLE collaborations (Bennett et al. 2002a;
Mao et al. 2002), could be a $\sim 100\,M_\odot$ black hole at a distance
of a few hundred parsecs, with the other interpretation being a $\sim
3\,M_\odot$ black hole in the Galactic bulge (Bennett et al. 2002b).  The
difficulty in being certain about the mass is that even with parallax
effects there are still degeneracies between lens mass and lens distance
that require modeling of the stellar distribution to make likelihood
estimates. For a given event, the lens mass depends inversely on the lens
distance, so there are large differences in the expected Bondi-Hoyle
accretion luminosity that may be resolved with radio or X-ray
observations.  In the meantime, it is interesting that these same
microlensing observations have placed strong limits on the fraction of
halo matter that can take the form of $1-30\,M_\odot$ black holes (Alcock
et al. 2001b).

\section{Formation Mechanisms for Intermediate-Mass Black Holes}

\begin{figure}[b!]
\hskip 0.3truein
\vbox{\hskip -0.6truein{
\psfig{file=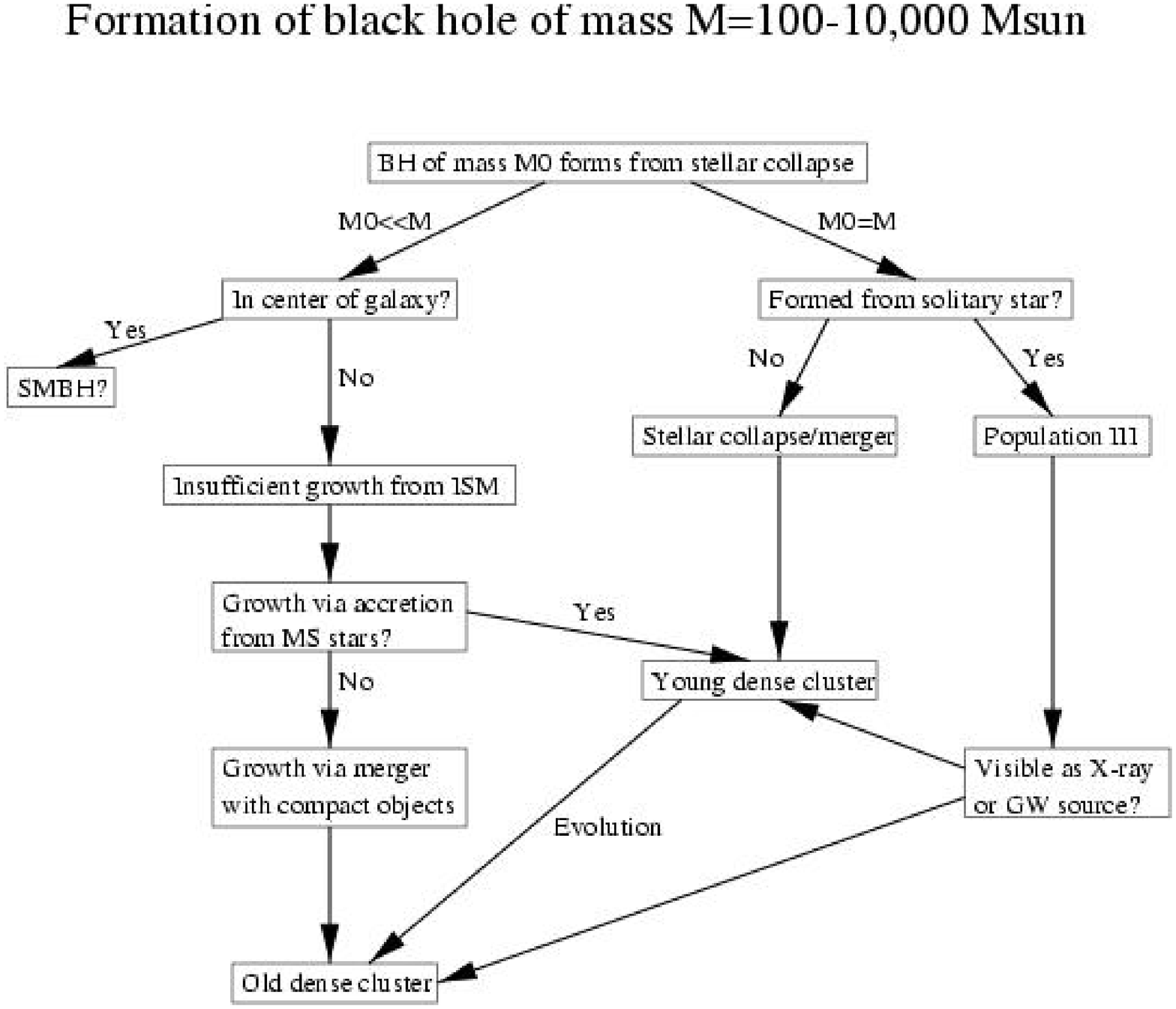,height=5.0truein,width=7.0truein,angle=270}}}
\caption{Summary of primary proposals for the formation of 
intermediate-mass black holes.  See text for details.}
\label{flowchart}
\end{figure}

Black holes in the $10^2-10^4\,M_\odot$ mass range are more massive than
the most massive stars that are forming in the current universe, although
there is uncertainty about the upper limit to current stellar masses.
General considerations of radiation forces on dust grains suggest that
past some mass, accretion will be suppressed (e.g., Larson \& Starrfield
1971; Khan 1974; Wolfire \& Cassinelli 1986, 1987; Jijina \& Adams 1996).
However, turbulent motion in the accreting gas can increase the limiting
mass significantly (McKee \& Tan 2003), and disk accretion could
circumvent many of the radiation force constraints (M. Wolfire, personal
communication). Observationally, the Pistol Star may have had an initial
mass $\sim 200\,M_\odot$ (Figer et al. 1998), and the most luminous stars
in NGC~604 could be well in excess of $120\,M_\odot$ (Bruhweiler, Miskey,
\& Neubig 2003).   Nonetheless, current thinking suggests that stars more
massive than $\sim 200\,M_\odot$ are unlikely to form in the current
universe, and even if they do then mass losses to winds and pulsations
will reduce the mass of any remnant black holes significantly (e.g., Fryer
1999). Therefore, if IMBHs exist they probably did not form recently from
core collapse. They either formed at some earlier time, or have
accumulated most of their mass since birth, or both.  The primary
formation mechanisms are summarized in Figure~\ref{flowchart}. In this
section we review some of the proposed mechanisms, in roughly
decreasing order of redshift. For a brief summary of formation mechanisms,
see also van der Marel (2003) and Miller (2003).

\subsection{Black holes in the very early universe}

An exotic possibility is that some class of black holes formed 
prior to big bang nucleosynthesis.  This has been explored as a
way to lock up a significant amount of matter in a non-baryonic 
form.  For example, during the QCD phase transition from quark matter
to nucleonic matter the equation of state of the universe was relatively
soft, and hence collapse may have proceeded with relative ease
(Jedamzik 1997, 1998; Niemeyer \& Jedamzik 1999).  However,
at this phase the horizon mass (i.e., the mass of causally connected
patches of the universe) was thought to be in the $\sim 1\,M_\odot$
range rather than $10^2-10^4\,M_\odot$ (Jedamzik 1997, 1998;
Niemeyer \& Jedamzik 1999), and since the
horizon mass increases with decreasing redshift, the intermediate-mass
black hole range would imply a transition at uncomfortably low energies.
In addition, the QCD black hole formation mechanism has been criticized
because it would require a perturbation spectrum that is strongly peaked
and finely tuned, in order to avoid constraints based on Hawking radiation
(Schwarz, Schmid, \& Widerin 1999).  This is therefore not the most
likely mechanism, but given our lack of knowledge of this phase of the
evolution of the universe, it bears keeping in mind.

\subsection{Population III stars}

A more promising early-universe origin for massive black holes is the first
generation of stars.  By definition, the so-called Population III stars
evolved in an environment with negligible metallicity, which in practice
appears to mean a metal fraction $Z\lta 10^{-5}Z_\odot$ (e.g., Abel et al.
1998; Bromm, Coppi, \& Larson 1999; Bromm et al. 2001; Abel, Bryan, \&
Norman 2000; Schneider et al. 2002; Nakamura \& Umemura 2002). In such an
environment, metal line cooling is absent and hence the temperature of
molecular clouds was higher than it is in the current universe.   The Jeans mass
scales as $T^{3/2}$, hence this suggests that the fragmentation mass and
thus the initial mass of stars may be significantly larger for Population
III stars than it is currently.  Moreover, a zero-metallicity star has
insignificant winds and weak pulsations (e.g., Fryer, Woosley, \& Heger
2001), so it loses comparatively little of its mass during its evolution.
A star with an initial mass between $100\,M_\odot$ and $250\,M_\odot$ is
believed to disrupt itself completely via a pair instability that leads to
explosive oxygen burning and leaves no remnant (e.g., Barkat, Rakavy, \&
Sack 1967; Woosley \& Weaver 1982, Bond, Arnett, \& Carr 1984; Carr, Bond,
\& Arnett 1984; Glatzel, El Eid, \& Fricke 1985; Woosley 1986; Heger \&
Woosley 2002), but above $250\,M_\odot$ the star is not disrupted and
instead is likely to collapse directly to a massive black hole, without an
explosion.

Madau \& Rees (2001) suggest that there may be $\sim 10^3-10^4$ such
black holes in a given galaxy.  Such a population would normally be
undetectable, because as isolated entities they would only be accreting
from the interstellar medium, which would generate little luminosity.
However, in an active star formation environment with many massive young
star clusters, isolated black holes could be captured gravitationally.
They would then sink to the center of the clusters, where they could
acquire a stellar companion and become active as X-ray sources (see also
Islam, Taylor, \& Silk 2003).  More
about this general scenario will be discussed below, but this mechanism
is a viable one for the ULXs.  Madau \&
Rees (2001) have suggested (see also Xu \& Ostriker 1994) that tens to
hundreds of Population III black holes will sink to the centers of their
host galaxies by dynamical friction, which may help in the assembly of
supermassive black holes and could be a significant source of
gravitational waves, detectable in the next decade at low frequencies by
space-based detectors such as the {\it Laser Interferometer Space
Antenna} (LISA; see, e.g., Danzmann 2000).

The lack of observational constraints on Population III stars means that
there is still substantial theoretical uncertainty about the mass range of
the first stars.  For example, cooling is likely to be dominated by
primordial H$_2$ (Matsuda, Sato, \& Takeda 1969; Yoneyama 1972; Hutchins
1976; Silk 1977; Yoshii \& Sabano 1980; Carlberg 1981; Lepp \& Shull 1984;
Palla, Salpeter, \& Stahler 1983; Yoshii \& Saio 1986; Shapiro \& Kang 1987;
Uehara et al. 1996; Haiman, Thoul, \& Loeb 1996; Nishi et al. 1998; Abel et
al. 1998; Omukai \& Nishi 1988; Couchman \& Rees 1986;  Bromm et al. 1999;
Abel et al. 2000; Nakamura \& Umemura 1999, 2001). If so, the temperature
can only get down to a few hundred Kelvin, leading to stellar masses
$>100\,M_\odot$.  However, in some circumstances (such as the presence of
dense shells or UV background radiation) the temperature may be lowered
enough for cooling by hydrogen-deuterium molecules to become effective
(Shapiro \& Kang 1987; Ferrara 1998; Susa \& Umemura 2000).  Since HD has a
nonzero permanent dipole moment (unlike H$_2$), the transitional temperature
is lower and hence cooling can proceed to temperatures of $T\lta 100$~K,
which produces stars of a few to tens of solar masses (e.g., Puy \& Signore
1996; Bougleux \& Galli 1997; Galli \& Palla 1998; Flower et al. 2000;
Nakamura \& Umemura 2002).   If this is common among the first generation of
stars then the initial black holes might still have been more massive than are
typically generated now, because of the lack of mass loss from winds and
pulsations, but the masses may not reach several hundred solar masses.  In
addition, it is not currently clear how many zero-metallicity stars can form
in a given galactic halo.  If most stars initially have masses greater than
$250\,M_\odot$ then there is little return of metals to the interstellar
medium and several such stars can form in a primordial environment.  If
instead many stars are formed below $250\,M_\odot$, then either
normal supernovae (below $100\,M_\odot$) or pair instability supernovae
(between $100\,M_\odot$ and $250\,M_\odot$) disperse the heavy elements and
rapidly raise the metal fraction above $10^{-4}Z_\odot$.  As this is an
active theoretical field, it is likely that in the next few years there will
be convergence on the expected physics and many of these issues will be
resolved.

\subsection{Dynamics of stellar clusters}

\subsubsection{Notes on Bondi accretion}

Suppose a black hole with current mass $M\sim 10^3\,M_\odot$ was
born with a much smaller mass.  A quick calculation shows that the
only way it could have grown to its current mass is if it spent
a significant amount of time in a dense stellar cluster, unless it somehow
spent billions of years in a supply of cool, dense gas.  Generically,
the mass could have been obtained via accretion from the interstellar
medium, accretion from a companion star, or mergers.  If the relative
velocity of the gas and black hole is dominated by the thermal velocity,
then accretion from
the interstellar medium proceeds at the Bondi-Hoyle rate
\begin{equation}
{\dot M}_{BH}\approx 10^{12}M_{100}^2\rho_{-24}T_6^{-3/2}\,{\rm g\ s}^{-1}
\end{equation}
(Bondi \& Hoyle 1944),
where the black hole mass is $M=100M_{100}\,M_\odot$, the density of the
interstellar medium is $\rho=10^{-24}\rho_{-24}$~g~cm$^{-3}$, and the
temperature of the interstellar medium is $T=10^6T_6$~K.  Any additional
velocity components, such as bulk velocity or turbulent velocity, decrease
this rate.  For the hot ISM, which comprises most of the volume of the ISM,
$T_6\approx 1$ and $\rho_{-24}\approx 10^{-2}-10^{-3}$ (e.g., Vogler \&
Pietsch 1999), so the mass accretion rate is $\approx
10^{9-10}M_{100}^2$~g~s$^{-1}$.  The e-folding time for mass increase is
then $(M/{\dot M})\approx 2\times 10^{17}M_{100}^{-1}$~yr for the hot ISM.
Even in a molecular cloud with $T\approx 100$~K, the accretion luminosity
itself will preheat the matter to $\sim 10^4$~K (e.g., Maloney, Hollenbach,
\& Tielens 1996; compare Blaes, Warren, \& Madau 1995 for accretion onto
neutron stars). Thus, even if $\rho_{-24}=100$, the e-folding time is
$\approx 6\times 10^{10}M_{100}^{-1}$~yr, much longer than a Hubble time
and thousands of times longer than  both the survival time of a molecular
cloud and the crossing time of a cloud for a black hole, even if the black
hole has a relative speed of only 1~km~s$^{-1}$. Therefore, unless the
black hole started out with a mass $M\approx 10^3\,M_\odot$, accretion from
the interstellar medium would add little to its mass.  This conclusion is
not necessarily valid in the centers of galaxies, where bar instabilities
and other processes may tend to funnel high-density, low-temperature gas
and foster growth of black holes. However, well off-center (as observed for
ULXs), there is no known process that would preferentially keep black holes
in an environment with low temperature and high density, as is needed to
promote growth by Bondi accretion.

\subsubsection{Merging and dynamics in clusters}

If instead a black hole grows by accretion from or merger with stars
or compact remnants, then because the stars or remnants themselves have
at most a few tens of solar masses, growth of several hundred solar
masses requires many encounters.  In galactic disks this has vanishing
probability because of the low number density of stars.  Therefore,
only in a dense stellar cluster would one have the required number of
objects with which the black hole could interact.  If the cluster has
a core relaxation time much less than the age of the universe (true for both
young stellar clusters and globular clusters) then substantial dynamical
evolution takes place and can assist in the growth of black holes.  In
this subsection we discuss general dynamical processes in clusters, and
in the following two subsections we will examine specific proposals for the
growth of intermediate-mass black holes in old (\S~4.3.3) or young
(\S~4.3.4) star clusters.

A cluster of $N$ stars in virial equilibrium with a crossing time
$t_{\rm cross}$ relaxes dynamically on a timescale
\begin{equation}
t_{\rm rel}={N\over{8\ln N}}t_{\rm cross}
\end{equation}
(e.g., Binney \& Tremaine 1987).  Over several relaxation times there are
a number of important trends of the evolution of the system. One is mass
segregation: more massive objects tend to sink to the center of the
cluster (on a timescale $(m_f/M)t_{\rm rel}$ for objects of mass $M$,
where $m_f$ is the average mass of a field star) while lighter objects
increase their scale height.  Therefore, generically, the more massive
stars or remnants in a cluster will tend to be found in the core. In
addition, because binaries act dynamically as a single object with the
combined mass of the two stars, binaries also have a tendency to sink
towards the core (e.g., Spitzer \& Mathieu 1980; see Elson et al. 1998
for observational evidence of binaries in a young stellar cluster).  
In a very young cluster (a few tens of millions of
years or younger), the most massive stars are still on the main sequence,
hence one expects O and B stars to be in the center.  In older clusters,
these massive stars have evolved off the main sequence and it is their
remnants (black holes or neutron stars) that are the most massive
objects. Thus, in older clusters one expects the core to be rich in black
holes, neutron stars, and binaries (see, e.g., Sigurdsson \& Phinney
1995; Shara \& Hurley 2001).

The presence of binaries adds crucial physics to the evolution of clusters.
Close interactions between binaries and single stars or other binaries
dominate the physics, hence these processes have been explored in a long
line of numerical experiments (e.g., Heggie 1975;  Hills 1975a,b; Hills \&
Fullerton 1980; Roos 1981; Fullerton \& Hills 1982;  Hut \& Bahcall 1983;
Hut 1983a,b; Hut \& Inagaki 1985; McMillan 1986; Rappaport, Putney, \&
Verbunt 1990; Mikkola \& Valtonen 1992; Heggie \& Hut 1993; Sigurdsson \&
Phinney 1993; Quinlan 1996; Portegies Zwart \& McMillan 2000). These
experiments have shown that in three-body interactions of a binary system
with a single object, hard binaries tend to get harder.  That is, for a tight
enough binary, the ultimate result of the interaction is usually that the binary
tightens further.  The resulting recoil adds to the kinetic energy budget
of the globular, which expands the system.  Such interactions are believed
to play the dominant role in preventing catastrophic core collapse in
globulars (originally suggested by He\'non 1961; see Ostriker 1985 and
Binney \& Tremaine 1987 for discussions).  Numerical experiments also show
that in a close three-body encounter of unequal masses, the tendency is
that the final binary is composed of the two most massive of the three
original stars (e.g., Sigurdsson \& Phinney 1993; Heggie, Hut, \& McMillan
1996).  As a result, even if a massive object is initially isolated,
interactions with binaries are likely to allow it to exchange in.  Thus,
massive compact objects in the cores of globulars are likely to be in
binaries.

Another trend evident in numerical simulations is that an individual
three-body encounter can be extremely complex, with hundreds or even
thousands of orbits required before the system resolves into a binary and a
single star.  During these many orbits, a pair of stars can come extremely
close to each other.  Newtonian simulations of three identical point masses
show (e.g., Hut 1984; McMillan 1986; Sigurdsson \& Phinney 1993) that the
probability that the closest encounter between two stars is $\epsilon a$ or
less (for initial semimajor axis $a$) during a binary-single interaction
scales as $\epsilon^{1/2}$.  Therefore, in a significant fraction of
encounters, two main-sequence stars in a young cluster may come close enough
to collide (Portegies Zwart \& McMillan 2002; see \S~4.3.4 below), or two
black holes in an old cluster may come close enough that gravitational
radiation is significant.

Finally, note that the large gravitational capture cross section of
stellar clusters means that even if an IMBH forms independently of a
young stellar cluster (as in the Population~III scenario), it can be
captured and sink to the core of a cluster, where it will form a
bright X-ray source if the most massive stars present are still on the
main sequence or giant branch.  The expected number of ULXs from this
mechanism is
\begin{equation}
N_{\rm ULX}=n_{\rm IMBH}N_{\rm cluster}\sigma_{\rm cluster}v_{\rm rel}T\; ,
\end{equation}
where $n_{\rm IMBH}$ is the number density of IMBH, $N_{\rm cluster}$ is the
number of super star clusters, $\sigma_{\rm cluster}$ is the cross section of
interaction of an IMBH with a cluster, $v_{\rm rel}$ is the relative velocity
at infinity of the IMBH, and $T$ is the lifetime of the young cluster.  
For a typical super star cluster, $M\sim 10^5\,M_\odot$ and $R_{\rm
cluster}\sim 10$~pc.  If the IMBH has had time to settle to nearly the
local standard of rest then it will have a slow speed and therefore be
only weakly hyperbolic relative to the cluster. For a relative speed of
$v_{\rm rel}=3$~km~s$^{-1}$ (which equals $3\times
10^{-6}$~pc~yr$^{-1}$, enough to travel 60~pc in $2\times 10^7$~yr), the
interaction is gravitationally focused and therefore $\sigma_{\rm
cluster}\approx \pi R_{\rm cluster} (2GM/v_{\rm rel}^2)$, or
$\sigma_{\rm cluster}\approx 3\times 10^3$~pc$^2$ for the chosen
numbers.

We now need to determine whether interaction with the cluster will
remove enough energy from the IMBH to make it bound to the cluster. From
the Chandrasekhar formula for dynamical friction (e.g., equation 7-18
from Binney and Tremaine), \begin{equation} d\ln{\bf v_M}/dt\approx
4\pi\ln\Lambda G^2\rho M/v_M^3\; . \end{equation} Here ${\bf v_M}$ is
the velocity of the massive object, $v_M$ is its magnitude, $\rho$ is
the average mass density of stars, and there is a proportionality factor
involving error functions that is roughly 0.4 for a weakly hyperbolic
encounters.

Assuming $\rho=10^5M_\odot/[(4\pi/3)(10~{\rm pc})^3]$, $M=10^3M_\odot$,
and $v_M=$10 km/s (because this is weakly hyperbolic), then for
$\ln\Lambda=10$ we have $d\ln{\bf v_M}/dt\approx 7\times
10^{-16}$s$^{-1}$. Therefore, a ``significant" change in velocity occurs
in a time of $t=1/7\times 10^{-16}~{\rm s}^{-1}=1.4\times 10^{15}$s.  At
$10^6$ cm/s, the distance traveled in that time is $1.4\times 10^{21}$
cm.  The diameter of the cluster is 20 pc, or $6\times 10^{19}$ cm.
Therefore, the fractional decrease in velocity is about 0.05.  The total
velocity is $(10^2+3^2)^{1/2}=10.4$ km/s, so a 5\% change will reduce
the speed below escape velocity and the IMBH will be captured by the
cluster.

Now consider the specific example of M82.  From the Hubble observations
of O'Connell et al. (1995), there are $\gta 100$ super star clusters in the
inner 350~pc of the galaxy.  If there are $N_{\rm IMBH}$ IMBHs in the same
volume, then $n_{\rm IMBH}\approx 5\times 10^{-7}(N_{\rm
IMBH}/100)$~pc$^{-3}$. If we consider $T=2\times 10^7$~yr, then the
expected number of ULXs by the capture mechanism is
\begin{equation}
\begin{array}{rl}
N_{\rm ULX}&\approx 5\times 10^{-7}(N_{\rm IMBH}/100)
{\rm pc}^{-3}\times
100\times 3\times 10^3{\rm pc}^2\times 3\times 10^{-6}{\rm pc~yr}^{-1}
\times 2\times 10^7{\rm yr}\\
&\approx 10(N_{\rm IMBH}/100)\; .\\
\end{array}
\end{equation}
If we assume the point sources in M82 (as listed in Table 1 of Matsumoto
et al. 2001) have a highly absorbed spectrum, similar to the ULX, this
implies there are 7 ULXs in M82, which is comparable to the expected
number of IMBHs.

\subsubsection{Gradual dynamical formation of IMBHs in globular clusters}

Three methods of forming IMBHs in globular clusters have been discussed:
\begin{enumerate}
\item  Merging of binaries that have a $\gta 50\,M_\odot$ black hole primary,

\item  Capture of a stellar-mass black hole in a high-eccentricity orbit
around an IMBH,

\item Tightening of a BH/BH binary by a Kozai resonance.
\end{enumerate}
We now discuss these in turn.

The first mechanism involves hardening of a binary by three-body
interactions, with the possibility of a merger due to gravitational
radiation if the binary is tightened enough. However, the recoil kick
experienced by both the binary and the interloper object is proportional to
the orbital velocity of the binary (e.g., Heggie 1975), so the kicks get
stronger as the binary hardens.  If the kick speed of the binary exceeds
the $\sim 50$~km~s$^{-1}$ escape speed from the core of a typical globular
(Webbink 1985) before the binary can merge via gravitational radiation,
then the final merger will happen well outside the globular (Portegies
Zwart \& McMillan 2000).  This may still be interesting for gravitational
radiation (Portegies Zwart \& McMillan 2000), but it would prevent
substantial growth of a black hole. Indeed, Kulkarni, Hut, \& McMillan
(1993), Sigurdsson \&  Hernquist (1993), Portegies Zwart and McMillan
(2002) find that three-body
interactions of  $10\,M_\odot$ black holes lead almost always to ejection
from globulars. In a young stellar cluster, where the core has not yet
contracted significantly, the escape velocity is much less, and ejection is
virtually inevitable.

If, however, the initial mass of the black hole is somewhat larger than
$10\,M_\odot$, it has greater inertia and might be able to stay in the cluster
long enough to grow significantly.  Miller \& Hamilton (2002a; see also
Taniguchi et al. 2000) show that black holes of initial mass
$M\gta 50\,M_\odot$ that interact with objects of typical mass
$M<10\,M_\odot$ have enough inertia that they merge by gravitational
radiation before ejection.  They can, therefore, grow in a dense cluster.
This process takes billions of years, and thus is not efficient enough to
form a black hole in a young stellar cluster. Many interactions are required
to harden a black hole binary to the point of merger, and a significant
fraction of these may eject the interloper stars, depleting the supply of
mass.  It is therefore important to evaluate how inefficient this process
is, which will allow an estimate of the maximum mass of a black hole that
can be grown this way.  Initial studies suggest that the high eccentricities
attained during three-body interactions allow merger by gravitational 
radiation when the binary still has a relatively large separation, and
hence the number of black holes kicked out during the hardening process
may be much smaller than estimated previously (G\"ultekin, Miller, \&
Hamilton 2003).

A second possibility, which could be more efficient for black holes of mass
$M>10^3\,M_\odot$, is direct capture by emission of gravitational radiation.
If two black holes, initially unbound with respect to each other, pass close
enough in a hyperbolic encounter, then emission of gravitational radiation
during the encounter may take away enough energy that the holes become bound.
From Quinlan \& Shapiro (1989), the effective capture cross section of a
compact object of mass $m$ by a large black hole of mass $M\gg m$ in such a
plunge is
\begin{equation}
\begin{array}{rl}
\sigma&= 2\pi\left(85\pi\over{6\sqrt{2}}\right)^{2/7}
{G^2m^{2/7}M^{12/7}\over{c^{10/7}v_\infty^{18/7}}}\\
&\approx 3\times 10^{28}
m_{10}^{2/7}M_{1000}^{12/7}v_6^{-18/7}\,{\rm cm}^2\; ,\\
\end{array}
\end{equation}
where $v_\infty=10^6v_6$~cm~s$^{-1}$ is the relative velocity
at infinity.  In cores with main sequence velocity dispersions
$v_{\rm ms}\approx 10^6$~cm~s$^{-1}$ and number densities
$n=10^6n_6$~pc$^{-3}$, the rate of such captures
is $\nu\approx 10^{-6}n_6m_{10}^{11/7}M_{1000}^{12/7}$~yr$^{-1}$
(Miller 2002).  This can be competitive with the rate of three-body 
encounters for large $M$, and can dominate the capture rate because
a single two-body encounter results in a merger, whereas many 
three-body encounters are required to harden the binary to the
point of merger (Miller \& Hamilton 2002a).  It is therefore possible
that the efficiency of black hole growth (defined as the change in
mass of the large black hole divided by the mass ejected) may increase
as the mass goes up.

The third channel for growth of an IMBH in a globular cluster has recently
been considered by Miller \& Hamilton (2002b), with consequences for
gravitational radiation discussed by Wen (2002).  A Newtonian
binary-single interaction of point masses cannot result in a stable
hierarchical triple.  However, a binary-binary interaction can, and the
sparse numerical results for this process suggest that a hierarchical
triple results from some 20-50\% of strong binary-binary interactions
(e.g., Mikkola 1984; McMillan, Hut, \& Makino 1991; Rasio, McMillan, \&
Hut 1995; Bacon, Sigurdsson, \& Davies 1996; Aarseth 2001). If the
resulting triple has a large inclination between the orbit of the outer
tertiary and the inner binary, then over many orbital periods the
eccentricity and inclination of the inner binary undergo a slow
oscillation called a Kozai resonance (Kozai 1962; Harrington 1968, 1974;
Lidov \& Ziglin 1976; Innanen et al. 1997; see Ford, Kozinsky, \& Rasio
2000 and Blaes, Lee, \& Socrates 2002 [which corrects an error in Ford et
al. 2000] for treatments to octupolar order). To lowest order, the net
result of the Kozai resonance is to change the eccentricity cyclicly from
its initial value to some potentially high value and back again, without
changing the semimajor axis of the inner binary.  If the maximum
eccentricity is sufficiently close to unity, gravitational radiation can
become important. Miller \& Hamilton (2002b) show that, even when
including post-Newtonian precession (which decreases the maximum
eccentricity), this process can often lead to merger by gravitational
radiation.  Since this does not produce any dynamical recoil, it may be a
way to build up massive black holes without ejections.  Given that even a
single $50\,M_\odot$ black hole can grow while staying in a cluster, from
three-body interactions, any path to their formation has potentially
important consequences.

There is, however, a possibility that even in Kozai mergers there can be a
significant kick.  The gravitational radiation emitted during the inspiral
of two black holes is not completely symmetric, because during one period
of revolution the orbit evolves.  Gravitational radiation therefore carries
away linear momentum, which imparts some velocity to the black holes (Peres
1962; Bekenstein 1973; Fitchett 1983; Fitchett \& Detweiler 1984; Redmount
\& Rees 1989; Wiseman 1992).   If this velocity exceeds $\sim
50$~km~s$^{-1}$, the merged system escapes from the globular.  The kick
velocity depends strongly on the radius of marginal stability ($v\sim
a_{ISCO}^{-4}$;  Fitchett 1983) and also on the mass ratio (by symmetry,
the kick vanishes for equal-mass nonrotating black holes; for Schwarzschild
holes the kick peaks at a mass ratio of 2.6; Fitchett 1983; Wiseman 1992).
The most recent post-Newtonian calculations suggest that the kick speed is
likely to be well below 50~km~s$^{-1}$ (Wiseman 1992), so this will not
eject black holes from clusters.  However, these calculations only address
the inspiral portion of coalescence, up to the innermost stable circular
orbit.  Current numerical and analytic calculations suggest that some 1-3\%
of the mass-energy of the system may be released in the final merger and
ringdown (Buonanno \& Damour 2000; Buonanno 2002; Baker et al. 2002),  so
it will be important to determine whether this phase produces substantial
asymmetry in radiation and thus may eject black holes.

\subsubsection{Rapid dynamical formation of IMBHs in young clusters}

In old clusters the most massive objects are stellar remnants. In
contrast, the most massive objects in clusters with ages  $\lta {\rm
few}\times 10^7$~yr are stars on the main sequence. These will sink to
the center of a cluster and perhaps decouple via the Spitzer instability
(Spitzer 1969), leading to a core collapse among only the most massive
stars.  The resulting high number density of massive main sequence stars,
combined with their large physical cross section, suggests that direct
collisions may be frequent.  If multiple such collisions occur for a
given object, a star with several hundred solar masses could be produced
(Begelman \& Rees 1978).  Ebisuzaki et al. (2001), Portegies Zwart \&
McMillan (2002), and G\"urkan, Freitag, \& Rasio (2003) have suggested
that this process can ultimately lead to the production of black holes of
masses $M\gta 10^2\,M_\odot$ in dense young stellar clusters, explaining
naturally the association of ULXs with star-forming regions in spiral
galaxies.  A similar process has been considered by Mouri \& Taniguchi
(2002b), and for a dense group of black holes within a stellar cluster by
Mouri \& Taniguchi (2002a).

There are a number of detailed questions with bearing on the domain of
applicability of this model.  The first has to do with time scales. If the
massive stars evolve off the main sequence before mass segregation is
effective, then the resulting mass loss in supernovae is enough to
decrease the central binding energy (and hence the central number density)
significantly, preventing rapid mergers (Portegies Zwart et al. 1999).
In addition, the remnants would be neutron stars or black holes and would
thus have negligible collision cross sections.  Portegies Zwart \&
McMillan (2002) estimate that if the half-mass relaxation time of a young
cluster is less than 25 million years, core collapse can occur and there
can be runaway growth of a central supermassive star.  If the relaxation
time is longer, mass loss from massive stars prevents core collapse.
However, even  relaxation times greater than 25 million years may lead to
core collapse if the initial star formation tends to place more massive
stars near the center of the cluster (for observations see Hillenbrand
1997; Hillenbrand \& Hartmann 1998; Kontizas et al. 1998; Sirianni et al.
2002, and for a theoretical perspective see Larson 1982; Bonnell et al.
2001). Even without pre-segregation of this type, young clusters are known
with central densities that may be high enough for rapid core collapse;
one example is R136 in the Large Magellanic cloud (Massey \& Hunter 1998).

The second question relates to the role of binaries.  In globular clusters,
formal core collapse (i.e., production of an infinite-density cusp at the
center) is prevented by the interactions of binaries.  The processes
discussed earlier in this section tend to tighten hard binaries, and if the
recoil speed is less than the escape speed from the cluster, the result is
injection of energy into the cluster, which tends to reduce the central
density.  Observations of young stars suggest that a large fraction of
them, particularly the high-mass stars, are in binaries. Therefore, one
possible outcome is that as the young binaries sink to the center of the
cluster, their interactions prevent the formation of a  high-density cusp,
which then prevents multiple collisions and the generation of a
supermassive star.  However, as pointed out by Hut \& Inagaki (1985),
McMillan (1986), and other authors, binary-single encounters of main
sequence stars tend to promote collisions because of resonant
interactions.  If binaries dominate the stellar fraction in the core,
binary-binary interactions can be especially important, and from Bacon et
al. (1996), the probability of a close approach is even larger than it is
for binary-single interactions (the numerical techniques used in this paper
actually underestimated the cross sections for $r_{\rm min}/a<0.01$, but
other results should be accurate [S. Sigurdsson, personal communication]).
There are therefore competing effects, between the tendency to heat the
center and the increased probability of collision in a single interaction.
Current simulations (Portegies \& Zwart 2002) do not include primordial
binaries, because of computational limitations (although progress is being
made; see Hurley, Tout, and Pols 2002; Hurley and Shara 2002;  Hut et al.
2003; Aarseth 2003), but further investigation is underway to determine the
regimes of parameter space in which each of these effects is most important.

The third question is what happens when two stars collide.  For the
low-velocity encounters expected in the center of a young cluster,
numerous studies have shown that a head-on collision is well-approximated
by a complete merger, with a loss of at most a few percent of the total
mass (e.g., Lai, Rasio, \& Shapiro 1993; Rasio \& Shapiro 1994, 1995).
However, the majority of collisions will not be head-on. There is thus
likely to be a tremendous amount of angular momentum in the merger
product.  Moreover, since the stars were originally  stably stratified
against convection and little entropy is likely to be generated in the
collision (because the collision velocity is much less than the sound
speed in almost all of the star), convection is not likely to transport
this angular momentum efficiently (Lombardi, Rasio, \& Shapiro 1996;  F.
Rasio, personal communication).  Other mechanisms, such as the
magnetorotational instability (Balbus \& Hawley 1991), are required to
dispose of the excess angular momentum.  There is, for example, the
possibility that for a significant time an expanding disk may exist
around this system, with consequences for further collisions and mass
accretion that are difficult to predict. Further research is necessary.
It may be that, since subsequent collisions probably occur at random
encounter angles, the overall angular momentum of the system is typically
small after many mergers and one can return to a ``sticky particle"
approximation.

Finally, assuming a collision leads to merger and production of a
higher-mass star, there is the question of how much mass is lost in the form
of winds and pulsational instabilities between collisions. Portegies Zwart
et al. (1999) find that although wind loss slows down the rate of growth of
supermassive stars it does not halt the growth entirely.  At stellar masses
$M\gta 100\,M_\odot$, there is much uncertainty about the relevant physical
processes.  Mass ejection from pulsations may limit the growth of stars
here, or other instabilities such as pair-production supernovae (Barkat et
al. 1967) may place limits at higher masses.  Current work suggests that
mass loss on the main sequence is unimportant, but that post main sequence
mass loss might expand the core significantly in some circumstances (e.g.,
G\"urkan, Freitag, \& Rasio 2003, in preparation). Understanding of these
processes will indicate how massive a black hole can arise from
main-sequence stellar collisions.

Monte Carlo analyses of young clusters seem promising as a way to include
most of the relevant physics without the computational limits imposed by
direct N-body integration; some initial results are reported by G\"urkan
et al. (2003). Even if the maximum mass turns out to be
much less than the masses of ULX sources or IMBH candidates in globulars,
it is possible that an initial $100\,M_\odot$ black hole formed in this
manner is a good seed for further growth by accretion of stars or, later,
by mergers with compact remnants.

\section{Alternate Explanations for ULXs}

The simplest arguments that ULXs are intermediate-mass black holes
are that their X-ray flux implies an isotropic luminosity well beyond
the Eddington luminosity of a $10\,M_\odot$ black hole, and that
their off-center locations in galaxies imply $M<10^6\,M_\odot$ from
dynamical friction arguments.  However, it is essential to recognize
that we currently have no {\it direct} measurements of the masses of
the black holes in ULXs.  As a result, we are forced to rely on
indirect observations and even more indirect theoretical arguments
to estimate the mass.  These are subject to significant error.  
It is therefore important to examine alternatives to these models,
as well as the arguments that have been raised against intermediate-mass
black holes as the engines of ULXs. 

\subsection{Beaming}

The most prominent alternative model has been a beaming model, in
which the source is a standard stellar-mass black hole with a jet or
relativistic beaming, which produces a flux in our direction that is
far in excess of the average flux and thus yields a misleading estimate
of the isotropic luminosity.  This idea was first proposed for ULXs by
Reynolds et al. (1997), and has recently been developed more
extensively by King et al. (2001), Markoff, Falcke, \& Fender (2001),
King (2002), and King \& Puchnarewicz (2002).  Koerding et al. (2001)
use population synthesis models for X-ray binaries to discuss the
statistical aspects of beaming models.  Butt, Romero, \& Torres (2003)
consider possible links between ULXs, microblazars, and EGRET sources,
and Foschini et al. (2003) list ULXs with power law spectra that may
be compatible with beamed emission.

The basic idea is simple: in many black hole sources, including the
microquasars and many AGN, there is known to be beaming, which may or
may not be relativistic depending on the source.
Indeed, jets are also known in other sources with accretion disks, such
as protostellar systems.  If we are in the beam of the jet, the flux can
be far beyond the average flux, either because of relativistic effects or
simply because of geometrical collimation. For example, Sikora (1981),
Madau (1988), and Misra \& Sriram (2003) have examined beaming from a
geometrically thick ``funnel" configuration, and found that the flux
along the axis of symmetry can be enhanced by a factor of tens compared
to the isotropic Eddington flux, depending on the opening angle of the
funnel and the height of its walls. There is also some geometric flux
enhancement expected in warped disk models, but this is more modest
(Maloney, Begelman, \& Pringle 1996; Pringle 1997).  Observationally,
evidence for such beaming may exist for some X-ray binaries: in their
review of black hole binaries, McClintock \& Remillard (2003) exhibit
several sources with fluxes that correspond to isotropic luminosities
of few$\times 10^{39}$~erg~s$^{-1}$.  The distances to these sources are
difficult to ascertain with precision, but combining the estimated luminosities
with estimated masses of the black holes indicate fluxes that can be a
few times the Eddington flux at the best derived distance (McClintock \&
Remillard 2003).

With such enhancement, the ULXs could be explained straightforwardly by
such beaming even if the black hole itself only has $M\sim 10\,M_\odot$.
King et al. (2001) propose that the ULXs in spiral galaxies are beamed
sources involving high-mass X-ray binaries during a phase of thermal
timescale mass transfer, and  King (2002) suggests that the ULXs in
elliptical galaxies are low-mass X-ray binaries that are microquasars with
their beam towards us (for a recent review of the relevant physics of
accreting compact binaries, including an application to ULXs, see King
2003).  King (2003) suggests that an optically thick outflow may even
account for the low temperatures reported from some XMM and Chandra
observations, but analysis also has to be done of whether this would imply
relativistic outflows.  For ULXs with fluxes corresponding to an isotropic
luminosity of less than $10^{40}$~erg~s$^{-1}$, the beaming factors
required are only a factor of a few.  For the most luminous ULXs, at $\sim
10^{41}$~erg~s$^{-1}$, a $10\,M_\odot$ black hole would have to have
emission beamed by a factor of $\approx 70$ to explain the flux while
remaining below the Eddington luminosity.

Beaming models have many advantages.  They explain the overall flux, they
are based on known source populations, and in spiral galaxies they explain
the association with star-forming regions.  However, there are a number of
challenges to such models.  For example, stellar-mass black hole candidates
such as Cyg~X-1 have power spectra that display significant variability up
to $\sim$100~Hz (e.g., Revnivtsev, Gilfanov, \& Churazov 2000).  In
contrast, ULXs have thus far not shown any rapid variability. Strohmayer \&
Mushotzky (2003) find no variability above the Poisson level at frequencies
greater than $\approx$0.1~Hz, despite being able to detect variability up
to 1~Hz.  If the beaming is relativistic, this would increase the
frequencies even more.  For the thermal timescale mass transfer suggested
by King et al. (2001), the mass transfer rate is well above Eddington,
which might have been expected to produce rapid transient obscuration of
the source (although without detailed models it is difficult to say this
with certainty).  As discussed in \S~2.7, it is especially difficult in the
beaming model to explain the 8.5\% rms QPO observed by Strohmayer \&
Mushotzky (2003) from an M82 ULX, because QPOs are thought to be a disk
phenomenon and in the beaming model the disk emission would make up only a
few percent of the total observed emission.

In addition, beamed black hole sources such as microquasars or blazars
often show relativistic outflow (e.g., the superluminal motion from
GRS~1915+105 detected in the radio by Mirabel \& Rodriguez 1994).  This
may be a consequence of super-Eddington fluxes caused by beaming, but the
theoretical basis for relativistic outflows is not sufficiently well
understood to make clear predictions.  Observationally, relativistic
beaming translates into an extremely flat $\nu F_\nu$ spectrum for
blazars; the typical $\nu F_\nu$ ratio of X-rays to $\sim 1-10$~GHz radio
waves is 10-1000 for blazars (Fossati et al. 1998).  In contrast, the few
ULXs that have been studied in radio usually show only upper limits to
radio emission, with even the detections giving X-ray to radio ratios of
$\sim 10^5-10^6$ (e.g., Kaaret et al. 2003; Neff et al. 2003).  For black
holes of stellar mass the radio cutoff frequency may be higher than it is
for supermassive black holes, but there is evidence that the X-ray to
optical ratio in ULXs is much larger than it is in beamed AGN as well (R.
Mushotzky, presentation at ``The Astrophysics of Gravitational Wave
Sources", 25 April 2003, College Park, MD; see also Zampieri et al. 2003).
Perhaps the emission mechanism is different or the different mass scale
produces different spectra, but this is an observation that needs to be
explained.

\subsection{Super-Eddington emission}

Even if the emission is quasi-isotropic, it could be that the luminosity is
well above the Eddington limit (Watarai, Mizuno, \& Mineshige 2001;
Begelman 2002).  Recall that, technically, the Eddington limit only applies
when (1)~the emission and accretion are quasi-isotropic, and (2)~the
dominant opacity is Thomson scattering. If magnetic fields are strong
enough, $B\gta 10^{13}$~G, Thomson scattering is suppressed and hence the
Eddington limit can be raised significantly, as is thought to be the case
for soft gamma-ray repeaters (e.g., Thompson \& Duncan 1995; Miller 1995).
Even for weaker fields $B\sim 10^{12}$~G, accretion in a column may be able
to provide enough anisotropy to allow super-Eddington emission (e.g., 
A0535-668, see Bradt \& McClintock 1983; Arons 1992 for a theoretical 
treatment).  However, black holes have no native magnetic fields and accretion
disks cannot attain the required fields, so this is not an option.  During
a supernova the accretion rate can be enormous, perhaps a few tenths of a
solar mass in a few seconds, and this is made possible because the
temperatures are great enough that neutrino emission dominates and
radiation forces are small as a result. But normal accretion of a star onto
a black hole misses these temperatures by three orders of magnitude, so
this is also unimportant.  The remaining possibility is anisotropy of
accretion and/or emission.

Begelman (2002) has pursued this line of thought and has conceived of a
potentially stable arrangement of matter in which radiation effects cause
accreting matter to clump.  These clumps are linked by weak magnetic
fields.  Radiation then moves primarily in the low-density medium between
the clumps, and this can in principle be stable.  Clumping of this type
has been observed in some numerical simulations (Turner et al. 2003). It
will be interesting to see this idea developed further, and in particular
to determine what conditions are necessary for it to occur.  It would be
especially useful to know what other observational
properties would distinguish such rapidly accreting black holes from
ordinary sub-Eddington accretors. Recent work by Ruszkowski \& Begelman
(2003) suggests that the total luminosity obtainable in these systems is
$\leq 10$ times the standard Eddington limit.  If further work confirms
this, it means that the most luminous of the ULXs might still require
higher masses than are usually considered for stellar-mass black holes.

\subsection{Motivation for stellar-mass models of ULXs}

The invocation of accreting stellar-mass black holes to explain ULXs
is reasonable because it involves a known class of sources and might
be able to match many of the observed properties of ultraluminous
X-ray sources.  In addition, there have been a number of concerns
about the viability of models involving intermediate-mass black holes.
Issues that have been raised include two observational and three
theoretical problems:

\begin{enumerate}

\item The inferred disk temperatures are too high for an IMBH.

\item The luminosity function of X-ray point sources shows no evidence
for a new component.

\item IMBHs cannot evolve in a binary.

\item IMBHs cannot grow in stellar clusters.

\item Separations of ULXs from young stellar clusters are inconsistent
with the IMBH model.

\end{enumerate}

We now examine these issues in turn.

{\it ULXs have high disk temperatures, therefore low mass.}---This
argument comes from the $kT\approx 1.4-1.8$~keV inner disk temperatures
inferred with a MCD model from ASCA data (Kotoku et al. 2000). Recalling
that in standard MCD fits, $kT\sim 1~{\rm keV}(M/10\,M_\odot)^{-1/4}$,
this would imply masses of just 1--2~$M_\odot$.  Therefore, if the
temperatures really are that high, it does indeed pose a problem for
models in which $M\sim 10^2-10^4\,M_\odot$.  However, as discussed in
\S~2, it is not clear whether the temperatures {\it are} that high.
None of the data collected with {\it Chandra} or {\it XMM} strongly
favor a multicolor disk model over alternate spectral forms such as
power laws or bremsstrahlung, even for sources for which ASCA data did
favor the MCD model (Smith \& Wilson 2003).  In addition, there are
several sources for which XMM and Chandra fits may suggest low disk
temperatures of a few tenths of a keV, consistent with black hole masses
in the hundreds to thousands of solar masses (e.g., Kaaret et al. 2003;
Miller et al. 2003a; Miller et al. 2003b).  It may be that the lack of
ASCA spectral coverage below 0.5~keV and/or its relatively large point
spread function introduce systematics in the fits.  This is an issue
that needs to be examined carefully, source by source, with {\it
Chandra} and {\it XMM}.  In addition, there are soft excesses in
narrow line Seyfert galaxies such as Ark~564 (Turner et al. 2001)
and Ton~S180 (Turner et al. 2002) that would imply high ``disk
temperatures", but these are clearly supermassive objects.

It is important to emphasize that continuum spectra can often be fit
well by a variety of models, which have very different interpretations.
With this in mind, one must exercise caution in linking spectral fits
to physical inferences about a system.  The recent work with {\it XMM}
and {\it Chandra} demonstrates only that previous claims of high
temperatures are not required, and should not be used as a strong
argument for low disk temperatures and hence high black hole masses.
Nonetheless, the current situation is fully consistent with the
intermediate-mass black hole hypothesis.

{\it The luminosity function of bright point sources has a constant slope,
hence a new population is excluded.}---The luminosity functions for X-ray
point sources in the Antennae (Zezas \& Fabbiano 2002) and NGC~4485/4490
(Roberts et al. 2002) can be fit by constant slope power laws.  As
mentioned by van der Marel (2003), one might expect that if a new
component is introduced at a high flux (e.g., intermediate-mass black
holes) the slope would change as a result.  Although potentially
interesting, this argument is not currently compelling, for two
reasons.

The most important is that the number of sources beyond the flux
corresponding to a $10\,M_\odot$ black hole at the Eddington luminosity
is so small that the error bars are huge.  For example, although the
luminosity functions from Zezas \& Fabbiano (2002) and Roberts et al.
(2002b) are broadly consistent with a single power law, they are also
consistent with a broken power law or many other forms.  For example, as
stated by Zezas \& Fabbiano (2002), there may be a break in the X-ray
luminosity function of the Antennae around $10^{39}$~erg~s$^{-1}$,
although it is not statistically significant.  There may be a similar
turnover in the Roberts et al. (2002b) data on NGC~4485/4490 at $\sim
10^{40}$~erg~s$^{-1}$, but in both cases the number of sources is far
too small to draw any conclusions.

Another problem is that if this argument is valid, {\it any}
change in the source population would be expected to change the slope in
the luminosity function.  For example, if at a certain flux level there is
a transition to beamed sources from the quasi-isotropic emission thought to
dominate the $L<10^{38}$~erg~s$^{-1}$ population, then one would expect a
break in the luminosity function because some fraction of the sources will
be missed.  Similarly, if the higher fluxes arise because of a new disk
geometry that allows super-Eddington luminosities, there is no reason to
expect that the power law slope will remain constant.  Perhaps at some
point the luminosity functions of point X-ray sources in many galaxies can
be combined to make a more statistically reliable statement
(for an initial analysis see Gilfanov et al. 2003), although care
would be needed in selecting a uniform sample of galaxies.

{\it A $M>100\,M_\odot$ black hole cannot evolve in a binary.}---This
applies to a scenario in which a very massive star (a few hundred solar
masses) in a binary with another star evolves to become an IMBH in a
binary. As argued by King et al. (2001), even if a such a massive
progenitor exists (which seems unlikely), Roche lobe overflow as opposed
to common-envelope evolution would demand an orbital period in excess of a
year.  Simple irradiated-disk theory (King \& Ritter 1998) may also imply
long-term luminosity behavior that is inconsistent with X-ray observations
(King et al. 2001).  Therefore, it appears highly improbable that a binary
system with an intermediate-mass black hole could have evolved in
isolation in the current universe.  However, as discussed in \S~4, a $\sim
10^2-10^4\,M_\odot$ black hole can {\it capture} a stellar companion in a
cluster.  Pending further investigation, this therefore appears to provide
a viable mechanism for the growth of IMBHs.

{\it A $M>100\,M_\odot$ black hole cannot grow in a cluster.}---A
separate argument is also discussed by King et al. (2001).  They consider
whether a stellar-mass black hole (some tens of solar masses) can, via
repeated dynamical encounters, grow to hundreds to thousands of solar
masses. Based on the simulations of Kulkarni, Hut, \& McMillan (1993) and
Sigurdsson \& Hernquist (1993), which involve binary-single interactions
of three $10\,M_\odot$ black holes,  dynamical interactions would eject
black holes, preventing growth.  However, there have now been scenarios
described in the literature that plausibly allow significant growth in
clusters.  As discussed in \S~4, Ebisuzaki et al. (2001), Portegies Zwart
\& McMillan (2002), and G\"urkan et al. (2003) have proposed that in
young stellar clusters there may be a significant number of direct
stellar collisions that produce a large star and an extra massive black
hole remnant after a supernova.  In older clusters, Miller \& Hamilton
(2002a) show that a black hole with initial mass $M\sim 50\,M_\odot$ can
stay and grow in a globular cluster, and that binary-binary interactions
can produce mergers without substantial dynamical kicks.  At this point,
therefore, growth of a black hole in a dense stellar cluster is still a
possibility.

Another potential concern is that asymmetric emission
of gravitational waves during coalescence could produce enough momentum flux to
kick black holes out of a globular or young stellar cluster.  As discussed
in \S~4, current evidence is that this would not happen, although numerical
calculations of the merger phase are needed for this to be definitive.

{\it A $M>100\,M_\odot$ black hole would not be separated from its
cluster.}---In their report of {\it Chandra} observations of ULXs in the
Antennae, Zezas \& Fabbiano (2002) point out that although there is an
excess of young stellar clusters near the sources, the X-ray sources are
usually displaced from the clusters, by a projected distance of 1-2"
(corresponding to 100-200~pc at a distance of 20~Mpc [for
$H_0=70$~km~s$^{-1}$~Mpc$^{-1}$]).  They argue that this rules strongly in
favor of lighter black holes, based on an assumed momentum kick during a
supernova that delivers a speed $v=245 (M_{\rm tot}/M_\odot)^{-1}$~km
s$^{-1}$, where $M_{\rm tot}$ is the mass of the black hole plus
companion. At 10~km~s$^{-1}$ a star will travel 300~pc in 30~Myr, which is
roughly the upper limit of the age of the young clusters, so Zezas \&
Fabbiano (2002) argue that the total mass can be no more than $\sim
20\,M_\odot$. If the supernova kick is greater, e.g., the $\sim
100$~km~s$^{-1}$ inferred for the $6\,M_\odot$ black hole in GRO~J1655
(Mirabel et al. 2002), then the upper limit is pushed to $\sim
60\,M_\odot$, but this is still well below the hundreds of solar masses
inferred from Eddington luminosity arguments.

Higher masses are allowed if there are other sources of velocity than the
initial supernova.  For example, as discussed in \S~4, three-body
interactions are expected to deliver a substantial set of kicks.  From
Table~3 of Maiz-Apellaniz (2001), the typical half-light radius of a
young super star cluster (similar to those in the Antennae) is 3--20~pc,
which for a mass of $10^5\,M_\odot$ implies an escape velocity of
7-15~km~s$^{-1}$ (young star clusters have not had as much time to
concentrate the densities in their cores, which is why the escape
velocity is less than that for globulars).  The question is then whether
three-body interactions can produce a $\sim 10$~km~s$^{-1}$ kick that
ejects the black holes from the clusters and produces the observed
separation.

From Quinlan (1996), in a three-body interaction of an interloper of
mass $m$ with a binary of total mass $M_{12}$, the recoil speed of the binary
after a three-body encounter is typically $v_{\rm kick}\approx 
0.85(m/M_{12})^{3/2}$ times the binary orbital velocity
$v_{\rm bin}=300(a/1~{\rm AU})^{-1/2}(M_{12}/100\,M_\odot)^{1/2}$~km~s$^{-1}$.
When $a\approx 10^{12}$~cm for typical masses, Roche lobe overflow
occurs for an early-type main-sequence star.  Therefore, if the required
kick is delivered for $a>10^{12}$~cm it is possible to have the observed
separation (larger separations will still undergo Roche lobe overflow
when the companion star evolves off the main sequence).  From the above
numbers, the kick velocity is
\begin{equation}
v_{\rm kick}\approx 40 (m/10\,M_\odot)^{3/2}(M/100\,M_\odot)^{-1}
(a/10^{12}~{\rm cm})^{-1/2}~{\rm km\ s}^{-1}\; .
\end{equation}
Therefore, if the most massive star typically interacting with the
black hole binary has $m=10\,M_\odot$, the required speed can be
attained for $M<400\,M_\odot$.  If $m=20\,M_\odot$, $M<1000\,M_\odot$
is allowed.  The time to dynamically evolve to this state is also
consistent; from the numbers in Miller \& Hamilton (2002a), 
$3\times 10^7$~yr suffices for the required orbital tightening
if $M<300\,M_\odot$ for $m=10\,M_\odot$, or if $M<1000\,M_\odot$ for
$m=20\,M_\odot$, assuming a central cluster number density of
$3\times 10^5$~pc$^{-3}$, which is comparable to the expected density of the
clusters in the Antennae (Zezas \& Fabbiano 2002).
The conclusion is that, whether intermediate-mass black holes are
formed in young clusters or formed elsewhere, dynamical three-body
kicks are consistent with the observed separation of ULXs from
their presumed parent clusters.
\bigskip\bigskip

Overall, the current state of the field is unsettled.  No definitive
observations exist for any single ULX, let alone the class of ULXs,
that rule out intermediate-mass black holes, or beaming, or
super-Eddington emission.  Observations of QPOs and Fe K$\alpha$ lines
such as in Strohmayer \& Mushotzky (2003) may come close to ruling out
beaming in individual sources, but there are still uncertainties in
modeling.  Only theoretical plausibility arguments exist, and the
weighting of them depends on the bias of the individual researcher.

What, then, could resolve the interpretation? Fundamentally, the current
disagreements exist because the crucial parameter ---  the mass --- has
not been measured observationally. Radial velocity measurements for any
source would be conclusive if they yield minimum masses greater than
$100\,M_\odot$.  If instead mass estimates of several of the brightest
sources consistently  produced $M\sim 10\,M_\odot$, models involving
stellar-mass black holes would be favored strongly.  Note that one cannot
make definitive statements about the entire class of ULXs, so it may be
that some ULXs conform to each of the models proposed.  In a slightly
more model-dependent way, observation of a ULX in an elliptical galaxy
would be important if the ULX is demonstrably far from any globular
cluster.  This is because all current IMBH models require the presence of
dense stellar clusters to feed the source.  An old (and therefore
low-mass) star in an elliptical would not be able to feed an IMBH at the
implied rates for long, therefore an observation of this kind would favor
beaming interpretations. It is also possible that long-term transient
behavior will distinguish between stellar-mass and more massive black
holes in binaries (Kalogera et al. 2003).  In \S~7 we discuss other
observations and theoretical calculations that will add important data.
We now consider some of the implications if intermediate-mass black holes
are common in the universe.

\section{Implications of Intermediate-Mass Black Holes}

\subsection{Formation of supermassive black holes}

If intermediate-mass black holes exist in abundance, then depending
on how they are formed they may have important implications for the
growth of black holes in the centers of galaxies.  For example, as
suggested by Madau \& Rees (2001), if Population~III stars form
$\sim 10^3\,M_\odot$ black holes then some of these may sink to the
centers of their host galaxies and form seeds for growth of supermassive
black holes by gas accretion.  It is also conceivable that some
supermassive black holes may have grown primarily by coalescence of
a number of intermediate-mass black holes.  This may be required if
some bright quasars exist at $z>10$, where the age of the universe
is $\sim 5\times 10^8$~yr, because at an Eddington mass
e-folding time of $4.4\times 10^7(\epsilon/0.1)$~yr, where 
$\epsilon\equiv L/({\dot M}c^2)$ is the accretion efficiency, it
would take $>8\times 10^8(\epsilon/0.1)$~yr to grow from $10\,M_\odot$ 
to $10^9\,M_\odot$.  If the accretion is in a disk with a constant sense 
of rotation this can be well in excess of a billion years, because the 
accretion efficiency increases for prograde orbits.  Coalescence of 
black holes is not subject to this limit, so rapid growth is possible.
However, it has been pointed out by Hughes \& Blandford (2002)
that any supermassive black hole with substantial rotation (as inferred,
e.g., from Fe~K$\alpha$ line profiles or jets) cannot have accumulated
its mass {\it primarily} by coalescence of small black holes on random
trajectories, because orbits in opposite directions lead to a net
small angular momentum.  It may, nonetheless, be possible to speed
up the process by building a $\sim 10^6\,M_\odot$ black hole by
coalescences, then accreting gas to the final current mass.

If the $M-\sigma$ relation derived for supermassive black holes in galaxies
(Ferrarese \& Merritt 2000; Gebhardt et al. 2000a; Merritt \& Ferrarese
2001a,b; Tremaine et al. 2002) also applies to globulars (Gebhardt et al.
2002; van der Marel et al. 2002; Gerssen et al. 2002), this has major
consequences for the origin of the correlation.  The central potentials of
globulars are extremely shallow compared to those of galaxies.  If,
nonetheless, black hole formation mechanisms are similar, this undoubtedly
is a major clue towards the origin of massive black holes, and possibly
towards whether the black holes have a fundamental influence in the
evolution of globulars and galaxies.

\subsection{Gravitational radiation sources}

Perhaps the most intriguing implication of intermediate-mass black
holes is their potential as gravitational radiation sources.  In
particular, if these objects are commonly found in the centers of
dense clusters, repeated interactions with other compact objects may
cause them to undergo tens to hundreds of coalescences per Hubble
time, many of which could be detected by ground-based or space-based
gravitational wave instruments.  These dynamical interactions produce
relatively high eccentricities, that could produce observable signals
of several post-Newtonian effects during inspiral.  Finally, intermediate-mass
black holes spiraling into supermassive black holes could generate
high signal to noise spectra with a large mass ratio that can be analyzed
to determine the three-dimensional structure of spacetime around a
rotating supermassive black hole.  We now discuss each of these 
possibilities in turn.

{\it Coalescence in clusters.}---As discussed in \S~4, a massive black hole
in a dense cluster will sink to the center, where it is likely to acquire a
binary companion in a short time (see also Benacquista 1999, 2000, 2002a,b;
Benacquista, Portegies Zwart, \& Rasio 2001).  Repeated interactions harden
the binary, leading ultimately to coalescence by gravitational radiation.
The overall rate of these events and their detectability depend on a number
of factors, including the fraction of star clusters that have a massive
black hole, the mass distribution of black holes, the typical mass of
objects that coalesce with the black holes, and the history of growth of the
black holes. Here we give a brief summary of one particular scenario that
illustrates many of these effects; for details see Miller (2002).

Suppose that the $M-\sigma$ relation found for supermassive
black holes in galaxies (see \S~3.1) also holds for globular clusters.
Then, from the number density of globulars and
the distribution of their velocity dispersions, we have an estimate
of the mass distribution of intermediate-mass black holes in globulars
and how common they are.  

A rough fit to the compilation of Pryor \& Meylan (1993) yields
a probability distribution for the velocity dispersion of Galactic
globular clusters of
\begin{equation}
P(\sigma)d\sigma\approx e^{-\sigma/4.3~{\rm km\ s}^{-1}}\left(
d\sigma/4.3~{\rm km\ s}^{-1}\right)\; .
\end{equation}
The current average number density of globulars is $n_{\rm GC}\approx
8h^3$~Mpc$^{-3}$, where $h\equiv H_0/100$~km~s$^{-1}$ is approximately
0.7 (Mould et al. 2000).  If the comoving density of globulars has
remained constant, then the locally measured density scales as $(1+z)^3$
and the number per steradian per unit redshift is (e.g., Peebles 1993, 
equation 13.61):
\begin{equation}
{dN\over{dz}}=n_{\rm GC}(H_0/c)^{-3}F_n(z)
\end{equation}
where $H_0$ is the Hubble constant and $F_n(z)$ is
\begin{equation}
F_n(z)=[H_0a_0r(z)]^2/E(z)\; ,
\end{equation}
where for a flat universe ($\Omega_m+\Omega_\Lambda=1$),
$E(z)=\left[\Omega_m(1+z)^3+\Omega_\Lambda\right]^{1/2}$ and
$H_0a_0r(z)=\int_0^z dz/E(z)$ (e.g., Peebles 1993).  For our
calculations we assume $\Omega_m=0.3$ and $\Omega_\Lambda=0.7$.
Roughly half of all globulars may have been destroyed in the last
$10^{10}$~yr by interactions with their host galaxies (Gnedin \&
Ostriker 1997; Gnedin, Lee, \& Ostriker 1999; Takahashi \& Portegies
Zwart 1998, 2000), so the number density at earlier epochs may have
been larger.  We will be conservative and not correct for this.

We now need to adopt a model for how the black holes in globular
clusters are built up.  Let us assume that such black holes are
born with an initial mass $M_{\rm init}$, and that they reach their
current masses (as given by the galactic $M-\sigma$ relation) via
accretion of compact objects of mass $\Delta M$.  Let us also assume
that this accretion occurs at a constant rate.  In reality, because
the effective interaction cross section of a black hole increases
with its mass, it is probable that the coalescence rate increases
with time.  Therefore, an assumption of a constant rate should
underestimate the detection rate somewhat, given that later coalescences
are at lower redshift and thus will give stronger signals.

The signals themselves are usually divided into the phases of inspiral,
merger, and ringdown.  Inspiral is the long stage from the initial
separation to the innermost stable circular orbit, merger is the
coalescence of the horizons, and ringdown is the final phase in which
the now single black hole settles into the Kerr spacetime.  Inspiral
and ringdown can be treated analytically (although the final stages of
inspiral need post-Newtonian treatments), but merger can only be
simulated numerically, meaning that reliable waveforms do not yet
exist for this phase.  The successive phases have higher and higher
frequencies.  For example, inspiral of a test particle into a nonrotating
black hole has a highest gravitational wave frequency (twice the orbital
frequency for a circular orbit), at the innermost stable circular
orbit, of $f_{\rm inspiral}=c^3/(6^{3/2}\pi GM)\approx 4.4\times 10^3
(M_\odot/M)$~Hz.  The characteristic highest frequency for merger and
ringdown is of order the light crossing time of the horizon, which
is $c^3/(GM)\approx 2\times 10^5 (M_\odot/M)$~Hz (with a factor less than
unity depending on the exact mode).  Therefore, ground-based detectors
(sensitive to frequencies $\sim 10-10^4$~Hz) could detect merger or ringdown
for most IMBH, but inspirals will be easiest to see for the lower-mass
members of this class.

\begin{figure}
\psfig{file=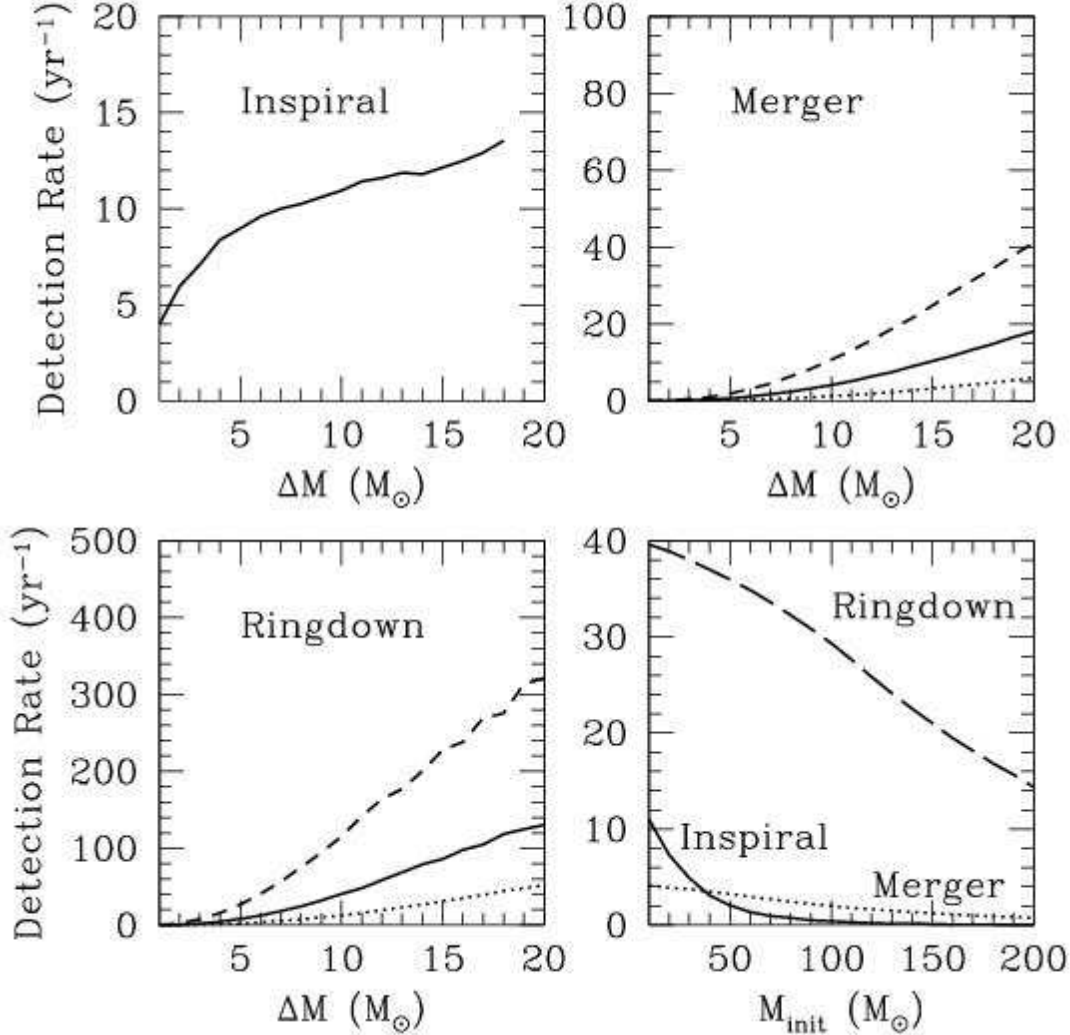,height=6.0truein,width=6.0truein}
\caption{Estimated rates of detection with the advanced LIGO detector
(in its standard configuration)
of stellar-mass compact objects falling into intermediate-mass black
holes.  The overall model, of black holes in globular clusters growing
by mergers, is described in detail in the text.  Top left: estimated
detection rate from the inspiral phase, as a function of the mass of
objects merging with the intermediate-mass black hole. In this and the
next two panels, the initial mass of the black hole is assumed to be
$10\,M_\odot$. Top right: estimated detection rate from the merger
phase, as a function of the mass of merging objects.  From bottom to
top, the curves assume an energy release during merger of
$0.005(4\mu/M)^2M$, $0.01(4\mu/M)^2M$, and $0.02(4\mu/M)^2M$,
where $M$ is the total mass of the binary and
$\mu$ is the reduced mass.  Bottom left: estimated detection rate
from the ringdown phase, as a function of the mass of the merging
objects.  Curves are the same efficiencies as in the top right panel.
Bottom right: detection rates in the inspiral, merger, and ringdown
phases as a function of the initial mass of the black hole, assuming
merger with $10\,M_\odot$ objects.  Here we assume energy releases
of $0.01(4\mu/M)^2M$ for both the merger and ringdown phases.}
\label{gravrad}
\end{figure}

The results of the globular cluster model are shown in
Figure~\ref{gravrad},  where the estimated rates in the inspiral, merger,
and ringdown phases are displayed as a function of the mass of objects
accreted, assuming an initial mass of $10\,M_\odot$ (first three panels),
and as a function of the initial mass, assuming accretion of $10\,M_\odot$
compact objects (bottom right panel). For the merger and ringdown panels
we show the rate as a function of accreted object mass for different
efficiencies $\epsilon$, where for a binary of reduced mass $\mu$ and
total mass $M$ we assume that during merger or ringdown a total energy of
$\epsilon M(4\mu/M)^2$ is released in gravitational waves.  Based on the
effective one-body calculations of Buonanno \& Damour (2000) and Buonanno
(2002), and the  numerical calculations of Baker et al. (2002), we
consider $\epsilon$= 0.005, 0.01, and 0.02 (from bottom to top in the two
panels).  In the bottom right panel we assume $\epsilon=0.01$. Consistent
with the estimates of Miller (2002), we find that some tens of events per
year involving intermediate-mass black holes are likely to be observable
with the planned advanced LIGO detector in its standard configuration,
with most of those coming during the merger or ringdown phases.  The noise
at low frequencies is decreased by lowering the laser power and working on
other non-fundamental design parameters, hence with tuning a larger number
of events could be detected.  Some mergers could be detected with mass
ratios greater than $\sim 40$, which might simplify numerical treatment.

{\it Long-term inspiral of black holes into IMBH.}---Whereas the
signals detected with ground-based interferometers will be during
the end of inspiral or later, space-based detectors such as the
{\it Laser Interferometer Space Antenna} (LISA) are sensitive to
lower frequencies, $\sim 10^{-4}-1$~Hz, and hence will observe the
longer-term inspiral of stellar-mass compact objects into 
intermediate-mass black holes.  As estimated by Miller (2002),
year-long LISA integrations are likely to detect several sources in
our Galactic globular clusters within a few million years of merger,
and several sources in globulars in the Virgo cluster of galaxies
within a few thousand years of merger.  The anticipated sources in Virgo are
particularly interesting, for several reasons.  First, the orbital
frequencies are expected to be $\gta 10^{-3}$~Hz, which is a range
relatively free from contamination by Galactic binary sources and most other
known sources of noise, hence the signals should be comparatively
clean.  Second, the expected dynamical interactions and evolutionary
paths of intermediate-mass black holes in clusters suggest that sources
from the Virgo cluster will have detectable pericenter precession,
orbital decay, and possibly Lense-Thirring precession (Miller 2002).
Along with the orbital period and eccentricity (which can be derived
from the waveform), this means that the gravitational wave signal alone
will suffice to determine the distance to the Virgo cluster; alternately,
assuming the distance as given, the orbital parameters will be 
overdetermined, leading to a strong self-consistency check of the
expressions for the different post-Newtonian effects (Miller 2002).

{\it Inspiral of IMBH into supermassive black holes.}---As suggested by
Madau \& Rees (2001; see also de Araujo et al. 2002; Cutler \&
Thorne 2002), it is
likely that some fraction of intermediate-mass black holes will
eventually merge with the supermassive black hole at the center of their
host galaxy.  These events have similarities to the process of capturing
stellar-mass compact objects by supermassive black holes, as considered
by Sigurdsson \& Rees (1997) and others. Such a merger would have a high
mass ratio and would therefore come close to the test particle limit
investigated in the context of stellar-mass black holes falling into
supermassive black holes (e.g., Hughes 2001).  These encounters are
thought to have great promise for mapping out the three-dimensional
spacetime near rotating black holes. If the small object is a $\sim
10^3\,M_\odot$ black hole instead of a $10\,M_\odot$ black hole, the
orbits are still nearly those of test particles but the signal to noise
ratio at a fixed luminosity distance is much greater.  Therefore, if
typically some tens of IMBH fall into a given supermassive black hole in
a Hubble time, and there are therefore tens of events per year of this
type that could be observed with LISA, this will provide extremely
precise measurements of the spacetime.

\section{Future Observations and Theory}

It should be clear from this review that there are numerous 
fundamental issues about intermediate-mass black holes that
have not reached consensus.  There is reason for optimism that
within a relatively short time, new observations and directed
theoretical calculations will give us new insights and possibly
resolve some of the current disputes.  In this section, we suggest
a number of such observations and theoretical developments.

{\it Radial velocity measurements.}---The only conventional astronomical way
that the mass of ULXs can be constrained rigorously is by radial velocity
measurements of stellar companions to ULXs.  For this to be definitive, one
needs to (1)~establish a periodicity for the X-rays, (2)~identify an
optical/UV/IR companion to the black hole that has the same orbital period
as the periodicity in the X-rays, and (3)~measure periodic Doppler shifts in
the spectrum of the companion.  This would produce a mass function, which is
a lower limit on the mass of the black hole.  If the mass function is
$M>100\,M_\odot$, the object is definitely an IMBH, proving their
existence.  If the mass function is $M\sim 10\,M_\odot$ or less, then in
principle the object could still be an IMBH (if, for example, the orbit of
the companion is close to face-on to us).  However, several examples of this
type would lead rapidly to a negligible probability that all the sources
were nearly face-on, and would therefore be strong evidence in favor of
stellar-mass models.  Even if no periodicity in X-rays is observed, a mass
function $M>100\,M_\odot$ would imply the existence of IMBHs, and one would
logically associate the IMBH with the nearest ULX.  For such searches it
would be best to concentrate on the highest-flux ULXs rather than ones close
to the fluxes of known  stellar-mass X-ray binaries.  Radial velocity
measurements will be challenging because optical counterparts are dim (e.g.,
Liu et al. 2002b) and because early-type stars have weak lines in the
optical.

{\it X-ray observations of ULX energy spectra.}---Now that high spectral
resolution observations are available from {\it Chandra} and {\it
XMM}, it is essential to revisit the spectra of the ULXs and
determine whether the conclusions reached with {\it ASCA} data still
apply.  If indeed there is significant optically thick disk emission
with temperatures $kT>$1~keV from high-flux ULXs, this argues in
favor of low-mass models such as the beaming scenarios.  If instead
the disk temperatures are usually significantly lower than those of
known stellar-mass black holes, or if alternate spectral models fit
at least as well as the multicolor disk models, these conclusions
need to be revised.  A systematic study of these spectra would be
valuable.

{\it X-ray timing observations.}---A potentially important type of
X-ray analysis that has not yet been pursued extensively is timing
measurements.  In stellar-mass black hole candidates, there are
characteristic bends in the power density spectrum that have been
suggested as guides to the masses of the compact objects (see Cropper
et al. 2003 for a promising application to NGC 4559 X-7).  The lower
the frequencies of these features, the higher the mass of the black
hole.  This may provide a strong discriminant between models of ULXs
involving stellar-mass black holes and those involving
intermediate-mass black holes.  This is particularly true for models
that involve relativistic beaming towards our line of sight, because
relativistic effects would further increase observed characteristic
frequencies.  Timing on scales of tens of milliseconds would be
ideal, but even archival {\it XMM} and {\it Chandra} data with
readout times of a few seconds may be valuable.  The benefits of this
approach have already been demonstrated by Strohmayer \& Mushotzky
(2003), whose discovery of quasi-periodic oscillations (along with an
Fe K$\alpha$ line) provide the strongest current evidence against
beaming in an individual ULX.

{\it Multiwavelength observations of ULXs.}---Many of the most important
observations of ULXs have come in the
optical.  These include associations with young stellar clusters (Zezas
et al. 2002) or globular clusters (Angelini et al. 2001);  the possible
identification of a stellar companion in M81 (Liu et al. 2002);  and the
optical nebular emission in Holmberg~II that may suggest quasi-isotropic
emission (Pakull \& Mirioni 2002).  There have also been a number of
nondetections in radio wavelengths that may hold clues to the emission
mechanism and whether or not many of these sources are beamed.
Focused multiwavelength observations of ULXs are likely to provide
additional breakthroughs.  For example, if more stellar companions to
ULXs are identified, it will determine whether the companions are always
early-type stars, or whether in ellipticals limits on companions can be
placed that suggest these are instead low-mass X-ray binaries. 

Multiwavelength observations are also important to evaluate the
broadband spectra of ULXs.  Models with beaming tend to predict
different spectra than models without, so UV/optical/IR/radio
observations of particularly bright ULXs will be important.  As more
systematic studies of this type are performed, a clearer picture will
emerge that can be compared to detailed spectral models.

Finally, multiwavelength observations may allow better estimates of
the overall luminosities of ULXs, rather than just their isotropic
equivalent luminosities.  Optical emission from nebulae surrounding
ULXs, as in the work of Pakull \& Mirioni (2002), may be one path.
It may also be possible to observe some molecular superbubbles to get
upper limits to the energy input of the ULX itself; in some cases, if
other energy sources can be neglected (e.g., early-type stars in a 
young cluster, or supernovae), an estimate of the energy output of the
ULX may be obtained.

{\it Kinematics of globular clusters.}---Just as observations of the
central regions of galaxies have provided strong evidence for
supermassive black holes and the $M-\sigma$ relation (Ferrarese \&
Merritt 2000; Gebhardt et al. 2000a; Merritt \& Ferrarese 2001a,b;
Tremaine et al. 2002), observations of the kinematics of the central
regions of globular clusters hold great promise for detection of massive
black holes or meaningful upper limits on their presence.  The current
observations, while promising, do not yet have enough statistical
significance to allow conclusions to be drawn. However, a sustained
observational campaign to look at globulars with high central velocity
dispersions (and thus, perhaps, high black hole masses) would be
extremely valuable, regardless of the result.  If no evidence for black
holes is found, this will shed light on the limitations of the formation
process and may guide research into the origin of the galactic $M-\sigma$
relation.  If black holes are found, this will be conclusive evidence for
intermediate-mass black holes in the universe and will immediately have
major implications for gravitational wave sources.

As discussed in \S~3, an even more promising avenue may be detection
of rotation in the cores of globulars.  If such rotation exists, and
if no explanation is found other than hardening of massive black binaries,
then this evidence indicates both that IMBHs exist in the cores of
globulars and that they are persistently and frequently merging with
stellar-mass black hole or other compact remnants (see \S~3).  Proper
motion data can augment radial velocity data significantly for both
this type of observation and for searches for cusps, and analysis of
such data is in progress (K. Gebhardt, personal communication; see also
Drukier and Bailyn 2003).

{\it Tasks for models of ULXs.}---In this review we have discussed three
different models for ULXs.  In one, these are stellar-mass black holes
that are beamed towards us, with luminosities less than the Eddington
luminosity.  In the second, the flux distribution is quasi-isotropic, but
the sources are stellar-mass black holes with luminosities many times the
Eddington luminosity.  In the third, the objects are intermediate-mass
black holes emitting quasi-isotropically below the Eddington luminosity.

All three models need to be able to address observational or theoretical
issues that have arisen in the last few years. If XMM and Chandra
observations can be modeled to determine uniquely the temperature of the
innermost portion of an optically thick disk, then a high temperature
will pose problems for models in which the black hole mass is
$M>10^2\,M_\odot$, whereas a low temperature would raise questions for
stellar-mass models.  There exist predictions for the multiwavelength
spectrum (radio to X-ray) for various beamed and unbeamed models, which
can be compared with data. Beaming models need to be able to account for
the lack of variability on short timescales, which might have been
expected by analogy to stellar-mass black holes in the Galaxy.
Super-Eddington models need to be explored more to determine the
conditions for such emission to occur in real disks, and to derive other
observational signatures that could be used to check the
interpretation.  In models with $M>10^2\,M_\odot$, more details need to
be understood about how such objects form.  For example, can
zero-metallicity (Population~III) stars really have hundreds of solar
masses?  If the objects are formed dynamically, then in detail can the
inferred masses be reached for the  known central densities and
populations of young or old clusters?  When partnered with observations,
there is much reason for optimism that many of these issues will be
resolved in the next few years.

{\it Gravitational waves.}---If intermediate-mass black holes exist in the
cores of many dense star clusters, one of their most exciting implications
is for the generation of gravitational waves.  In addition to the
dynamical simulations discussed above, this may pose new challenges for
numerical simulations of mergers.  A stellar-mass compact object merging
with an intermediate-mass black hole does so with a mass ratio of $\sim
10:1$ to $\sim 1000:1$.  Such mass ratios, particularly at the low end of
this range, are not large enough that the less massive object can be
considered as a test particle to high precision. However, the interactions
are likely to be less complicated to simulate than equal-mass mergers.
This may therefore provide a bridge calculation on which to test numerical
codes.

Finally, we note that debate on the nature of IMBHs will undoubtedly
continue until rigorous measurements of the masses of IMBHs are possible.
If radial velocity measurements are impractical because of the low fluxes
from stars at the distance of ULXs, then the only other way to measure
the mass beyond doubt is to observe the gravitational waves from the
inspiral of a compact object into an IMBH.  If so, then IMBHs are,
remarkably, one of the small set of objects in the universe whose proof
of existence requires detection of their gravitational radiation emission.

\acknowledgements 
We have had many fruitful discussions with Monica Colpi, Riccardo DeSalvo,
Karl Gebhardt, Doug Hamilton, Andrew King, Jon Miller, Fred Rasio, Chris
Reynolds, Tim  Roberts, Steinn Sigurdsson, Yuichi Terashima, Kip Thorne,
Roeland van der  Marel, and  Andrew Wilson. We are also grateful for the
figures supplied by Lorella Angelini, Franz Bauer, Pepi Fabbiano, Gordon
Garmire, Karl Gebhardt, Joris Gerssen, Ji-Feng Liu, Jon Miller, Manfred
Pakull, and Tim Roberts. This work was supported in part by NSF grant
AST~0098436 and NASA  grant NAG 5-13229 at the University of Maryland, and
NASA grant NAG 5-11670 at Johns Hopkins University.

\end{document}